\documentclass[twocolumn]{aastex631}

\usepackage[T1]{fontenc}

\received{2025 January 28}
\revised{2025 March 31}
\accepted{2025 April 17}

\shorttitle{HAT-P-67~b New Density}
\shortauthors{Wang et al.}

\begin{document}

\title{A Revised Density Estimate for the Largest Known Exoplanet, HAT-P-67~b} 

\author[0000-0003-3092-4418]{Gavin Wang}
\correspondingauthor{Gavin Wang}
\email{gwang59@jhu.edu}
\affiliation{William H. Miller III Department of Physics \& Astronomy, Johns Hopkins University, 3400 N. Charles Street, Baltimore, MD, USA}

\author[0000-0001-6396-8439]{William O. Balmer}
\altaffiliation{Johns Hopkins University George Owen Fellow}
\affiliation{William H. Miller III Department of Physics \& Astronomy, Johns Hopkins University, 3400 N. Charles Street, Baltimore, MD, USA}
\affiliation{Space Telescope Science Institute, 3700 San Martin Dr, Baltimore, MD, USA}

\author[0000-0003-3818-408X]{Laurent Pueyo}
\affiliation{Space Telescope Science Institute, 3700 San Martin Dr, Baltimore, MD, USA}

\author[0000-0002-5113-8558]{Daniel Thorngren}
\affiliation{William H. Miller III Department of Physics \& Astronomy, Johns Hopkins University, 3400 N. Charles Street, Baltimore, MD, USA}

\author[0000-0001-8510-7365]{Stephen P.\ Schmidt}
\altaffiliation{NSF Graduate Research Fellow}
\affiliation{William H. Miller III Department of Physics \& Astronomy, Johns Hopkins University, 3400 N. Charles Street, Baltimore, MD, USA}

\author[0000-0002-6379-3816]{Le-Chris Wang}
\affiliation{William H. Miller III Department of Physics \& Astronomy, Johns Hopkins University, 3400 N. Charles Street, Baltimore, MD, USA}

\author[0000-0001-5761-6779]{Kevin C.\ Schlaufman}
\affiliation{William H. Miller III Department of Physics \& Astronomy, Johns Hopkins University, 3400 N. Charles Street, Baltimore, MD, USA}

\author[0000-0001-7409-5688]{Gu\dh mundur Stef\'ansson} 
\affiliation{Anton Pannekoek Institute for Astronomy, University of Amsterdam, Science Park 904, 1098 XH Amsterdam, The Netherlands} 

\author[0000-0003-4408-0463]{Zafar Rustamkulov}
\affiliation{Morton K. Blaustein Department of Earth \& Planetary Sciences, Johns Hopkins University, Baltimore, MD, USA}

\author[0000-0001-6050-7645]{David K. Sing}
\affiliation{William H. Miller III Department of Physics \& Astronomy, Johns Hopkins University, 3400 N. Charles Street, Baltimore, MD, USA}
\affiliation{Morton K. Blaustein Department of Earth \& Planetary Sciences, Johns Hopkins University, Baltimore, MD, USA}

\begin{abstract}

Low-density ($\rho < 0.1 \mathrm{~g} \mathrm{~cm}^{-3}$) hot Saturns are expected to quickly ($<100$~Myr) lose their atmospheres due to stellar irradiation, explaining their rarity. HAT-P-67~b seems to be an exception, with $\rho < 0.09 \mathrm{~g} \mathrm{~cm}^{-3}$ and maintaining its atmosphere to well after 1 Gyr. We present a photometric and spectroscopic follow-up of HAT-P-67~b to determine how it avoided mass loss. HAT-P-67~b orbits a $V=10.1$ evolved F-type star in a $4.81$ day orbit. We present new radial velocity observations of the system from the NEID spectrograph on the WIYN 3.5m Telescope from a follow-up campaign robust to stellar activity. We characterize the activity using photometry and activity indicators, revealing a stellar rotation period ($5.40\pm0.09~\mathrm{d}$) near HAT-P-67~b's orbital period. We mitigate the stellar activity using a constrained quasi-periodic Gaussian process through a joint fit of archival ground-based photometry, TESS photometry, and our NEID observations, obtaining a planetary mass of $M_p = 0.45 \pm 0.15$~M\textsubscript{J}. Combined with a radius measurement of $R_p=2.140 \pm 0.025$~R\textsubscript{J}, this yields a density of $\rho_p = 0.061^{+0.020}_{-0.021} \mathrm{~g} \mathrm{~cm}^{-3}$, making HAT-P-67~b the second lowest-density hot giant known to date. We find the recent evolution of the host star caused mass loss for HAT-P-67~b to only recently occur. The planet will be tidally disrupted/engulfed in $\sim 150-500$~Myr, shortly after losing its atmosphere. With rapid atmospheric mass loss, a large, helium leading tail, and upcoming observations with the Hubble Space Telescope, HAT-P-67~b is an exceptional target for future studies, for which an updated mass measurement provides important context. 

\end{abstract}


\defcitealias{Zhou_2017}{Z17}

\section{Introduction} \label{sec:intro}

The most common method of determining exoplanet masses is the radial velocity (RV) method, which measures the Doppler shifts induced on the host star’s spectrum by the motion of the star-planet system about its common barycenter \citep[e.g., ][]{struve1952, 2010exoplanets, perryman2018exoplanet}. Given the mass of the host and the amplitude of the Doppler signal, a planetary mass, depending on the inclination of the planet's orbit with respect to the line of sight, can be inferred. Despite an order of magnitude improvement in precision \citep[e.g., ][]{Mahadevan2012, schwab2016, kpf2016, pepe2021} since the first RV detection of an exoplanet \citep{1995Natur.378..355M}, the impact of stellar activity on the detectable stellar semi-amplitude has confounded the detection and characterization of planets with this technique \citep{SuarezMascareno2022, Blunt_2023}. Variations in the star's photosphere, magnetic field, or circumstellar environment impart stochastic noise onto the spectrum of the host star, which can influence Doppler measurements \citep{Saar_1997, Robertson_2015, Gupta_2021, Tran_2024}. Disentangling and revealing the planetary signal from this stochastic noise requires detailed analysis of the data \citep[e.g.,][]{Nava_2020, Gan_2020, Blunt_2023}. 

As of January 2025, HAT-P-67~b is reported to be the second lowest-density transiting hot giant planet known by the NASA Exoplanet Archive,\footnote{\url{https://exoplanetarchive.ipac.caltech.edu/cgi-bin/TblView/nph-tblView?app=ExoTbls&config=PS}} behind WASP-193~b \citep{Barkaoui2024}. With a radius of $2.1$ R\textsubscript{J} and a mass of only $0.34^{+0.25}_{-0.19}$ M\textsubscript{J} \citep[hereafter \citetalias{Zhou_2017}]{Zhou_2017}, HAT-P-67~b exists in an exceptional density regime of $\rho<0.1~\mathrm{~g} \mathrm{~cm}^{-3}$. Recent atmospheric transmission studies by \citet{Bello_Arufe_2023}, \citet{gully_2023}, and \citet{Sic24} have found evidence for a large, leading helium tail and a high mass loss rate, which indicate that the planet is evaporating due to the incident radiation from its host star. Given the planet's equilibrium temperature \citepalias[$1900$~K;][]{Zhou_2017}, low density, and large scale height ($\sim 800$~km), it is an exceptional candidate for transmission spectroscopy to probe limb-to-limb variations, atmospheric dynamics, cloud condensation, and more. A mass uncertainty of 75\%, however, inhibits an accurate determination of the planet's atmospheric properties \citep{Batalha2019}.

\citet{Thorngren_2023} demonstrates, however, that low-density hot Saturns are expected to quickly lose their outer gas layers through runaway mass loss driven by extreme ultraviolet (XUV) irradiation from the host star. This runaway effect reproduces the observed trends in mass-radius space for the current population of hot-Saturns, namely the paucity of gas giants with densities below $0.1\mathrm{~g} \mathrm{~cm}^{-3}$. The mass measurement of HAT-P-67~b from \citetalias{Zhou_2017} places the planet significantly outside this regime, such that it is expected to have lost all of its envelope by its current age.

This observation has two possible explanations. First, there has been a recent change in the system, either inward migration or the evolution of the star into the subgiant branch (such that the XUV flux from the host star has not had sufficient time to remove its atmosphere). Second, stellar variability could have misled the mass measurement, and the planet is instead more massive. The host star HAT-P-67 has a high $v\sin i_{*}$ of $\sim30 \mathrm{~km/s}$ \citepalias{Zhou_2017} and has had signs of activity yielding large changes in radial velocity \citep{Sic24}.

In this paper, we differentiate between these two hypotheses through a detailed investigation of the star’s radial velocities. We confirm a Saturn-like mass for the planet at higher statistical significance with additional radial velocity observations from the NEID spectrograph on the WIYN 3.5m Telescope at Kitt Peak National Observatory (KPNO),\footnote{The WIYN Observatory is a joint facility of the NSF's National Optical-Infrared Astronomy Research Laboratory, Indiana University, the University of Wisconsin-Madison, Pennsylvania State University, Purdue University and Princeton University.} preserving a $<0.1\mathrm{~g} \mathrm{~cm}^{-3}$ density. This provides evidence in favor of a recent change in the system's state, the first hypothesis presented above. 

The remainder of this paper is organized as follows. \S\ref{sec:data} describes our photometric and spectroscopic data, as well as the modeling procedures we implement for each dataset. In \S\ref{sec:analysis} we perform a joint modeling of all data and present updated planetary parameters for HAT-P-67~b, in particular a revised mass of $M_p = 0.45 \pm 0.15$~M\textsubscript{J}, radius of $R_p = 2.140 \pm 0.025$~R\textsubscript{J}, and a density of $\rho_p = 0.061^{+0.020}_{-0.021}~\mathrm{~g} \mathrm{~cm}^{-3}$. \S\ref{sec:discussion} discusses the theoretical implications of our results and the fate of the planet. We end with a conclusion in \S\ref{sec:conclusion}. 

\section{Data} \label{sec:data}

\subsection{NEID Radial Velocities} \label{subsec:neid}

NEID \citep{schwab2016} is a high-resolution ($R=100,000$) temperature-stabilized \citep{Stefansson_2016, Robertson_2019} spectrograph on the 3.5-meter WIYN telescope at Kitt Peak National Observatory. NEID covers a broad wavelength range from 380 to 930~nm. At the time of our observations (2022 to 2023), HAT-P-67~b had already been discovered with a precise ephemeris, but had an uncertain mass measurement of $0.34^{+0.25}_{-0.19}$~M\textsubscript{J} \citepalias{Zhou_2017}. As the goal of our RV follow-up was to determine the mass with higher precision, we preferentially targeted the RV quadrature phases ($-0.25$ and $+0.25$, with phase 0 being when the planet transits), when the RV amplitude would be expected to be greatest. Our observing strategy was to take 3 consecutive exposures of 600 seconds each night the planet was at or near quadrature. This strategy helps to average over RV variations driven by high frequency stellar oscillations. Given the estimated RV jitter, $\sigma_\mathrm{jit}=59~\mathrm{m/s}$ (\citetalias{Zhou_2017}, their Table 6), after the three observations are added in quadrature, this strategy was intended to achieve an uncertainty of $\sim 64~\mathrm{m/s}$. We note that the RV jitter is of comparable amplitude to the planetary signal, which is larger than for solar-type stars \citep[e.g., jitter-to-planet amplitude ratio of $\sim 0.1$ for 51 Peg;][]{1995Natur.378..355M} but not the largest \citep[e.g., ratio of $5-8$ for V1298 Tau;][]{SuarezMascareno2022}. 

Observations of HAT-P-67~b were initially planned to be taken from May through July of 2022 but were cut short due to the Contreras Fire on 2022 June 12. Only 27 spectra (3 spectra per night for 9 nights) were taken between 2022 May 24 and 2022 June 12 before the site had to be shut down. Due to the target's visibility from the site (February through July), observations resumed the following year and an additional 48 spectra were taken from 2023 February 27 to 2023 July 5. In total, this amounts to 75 spectra, with the majority being taken after the fire. During the fire, the instrument was put through a vacuum cycle, resulting in an RV offset e.g., due to subtle changes in the instrument point spread function. Thus, in our analysis, we treat the RV data before and after the fire as separate independent instruments, assigning them independent RV offsets and jitter values. 

\begin{deluxetable*}{cccccc}
\tablecaption{Radial velocity data from NEID. The full table is available online in machine-readable form. } \label{tab:rvs}
\tablehead{\colhead{BJD\textsubscript{TDB}} & \colhead{RV (m/s)} & \colhead{$\sigma$\textsubscript{RV} (m/s)} & \colhead{Instrument$^{\text{a}}$}} 
\startdata
$2459723.87646187$ &$78.555$ &$52.491$  &NEID1\\
$2459723.88356815$ & $-73.380$ & $40.301$ &NEID1\\
$2459723.89063798$ &$-75.033$ &$37.503$  &NEID1\\
$2459725.90326585$ &$-8.646$ &$16.371$  &NEID1\\
$2459725.91048804$ &$-45.845$ &$16.097$ &NEID1
\enddata
\begin{itemize}
    \item[] $^{\text{a}}$ NEID1 refers to data collected before the June 2022 Contreras Fire; NEID2 refers to data collected after 
\end{itemize}
\end{deluxetable*}

\subsubsection{SERVAL Pipeline}
The NEID spectra were extracted using the NEID Data Reduction Pipeline (DRP)\footnote{\url{https://neid.ipac.caltech.edu/docs/NEID-DRP/}} and retrieved from the NEID archive.\footnote{\url{https://neid.ipac.caltech.edu/search.php}} To extract RVs, we used a modified version of the Spectrum Radial Velocity Analyzer \citep[SERVAL;][]{Zechmeister_2018} pipeline following \citet{Stefansson_2022}, which uses as-observed spectral templates to extract RVs from the data. This resulted in median RV uncertainties of 26~m/s in the unbinned data. For comparison, the RV uncertainty reported by the NEID DRP, which are based on cross-correlation functions, were 124~m/s and showed large RV variations of $>500$~m/s, oftentimes over a length of just a few days, and such variations were not seen in the HIRES dataset. We believe these excursions in the DRP RVs are likely due to the broad CCFs. Instead, the SERVAL pipeline self-consistently accounts for the RV uncertainties in the broadened as-observed templates. Given that our data spanned the instrument vacuum cycle due to the Contreras fire, we experimented extracting RVs from templates that spanned the whole data stream along with independently extracting RVs from templates generated using data before and after the fire. In both cases the RVs were equivalent. As such, we elected to use the extraction that generated a template using all of the data.

\begin{figure}[ht!]
    \centering
    \includegraphics[width=\columnwidth]{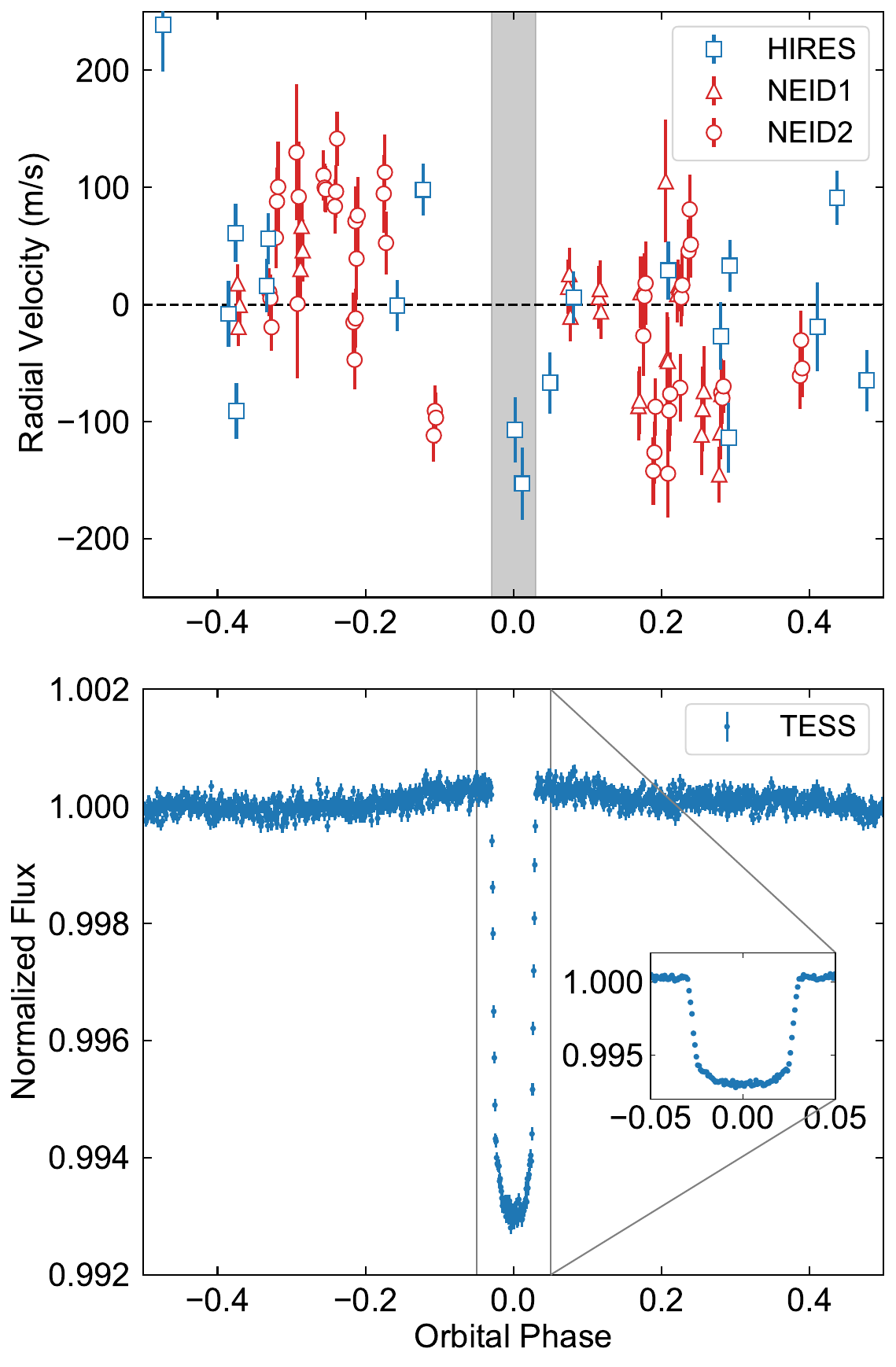}
    \caption{\textbf{Top:} Zero-pointed and phased RVs including both the HIRES (blue squares) and NEID (red triangles and circles) data. The two HIRES datapoints which fall in the shaded region are affected by the RM effect and thus excluded. The two datasets exhibit similar variational amplitude consistent with the upper bound from \citetalias{Zhou_2017}. \textbf{Bottom:} Phase-folded light curve from TESS with bins of 100 points, demonstrating the existence of HAT-P-67~b.}
    \label{fig:with-keck}
\end{figure}

\subsection{HIRES Radial Velocities} \label{subsec:hires}

The High Resolution Echelle Spectrometer \citep[HIRES;][]{Vogt_1994} is a high-resolution ($R = 55,000$) spectrometer on the 10 m Keck Observatory at the Mauna Kea Observatories. High-precision radial velocities from HIRES of HAT-P-67 were published by \citetalias{Zhou_2017} (see their Table 3). A total of 19 spectra were taken between July 2009 and March 2012. \citetalias{Zhou_2017} did not detect an obvious orbit for the planet, and thus provided an upper constraint on the RV semi-amplitude based on the data. We are able to replicate the results in that work through a \texttt{juliet} \citep{Espinoza2019} fit of all 19 measurements using wide priors and without the use of a Gaussian process. For our joint analyses, we discard the two data points which are collected at phase $0.002$ and $0.012$, which are when the planet is transiting. Due to the rapid spin of the host star, the Rossiter-McLaughlin (RM) effect is non-negligible \citep{Rossiter_1924, McLaughlin_1924}, and indeed \citet{Sic24} finds a $200~\mathrm{m/s}$ semi-amplitude RM effect for HAT-P-67~b. All radial velocity measurements phased to the planet period are plotted in Figure \ref{fig:with-keck}. 

Due to the low mass of the companion, the sinusoidal signal in the radial velocities is not immediately obvious, although the average radial velocity for phases $>0$ and $<0$ are noticeably different. To briefly confirm the information content of the observations, we fit all of the radial velocity data while keeping the period fixed but the phase ($t_0$) free, and we find we are able to reproduce a phase consistent with that from \citetalias{Zhou_2017}.

\subsection{HATNet Photometry} \label{subsec:hatnet}

The Hungarian-made Automated Telescope Network \citep[HATNet; ][]{Bakos_2004} monitored HAT-P-67 over a period of 7 months each in 2005 and 2008. This light curve data is publicly available in three forms: raw, External Parameter Decorrelation (EPD), and Trend Filtering Algorithm (TFA) magnitudes. Upon converting these magnitudes to relative fluxes and fitting with a single-planet transit model at HAT-P-67~b's ephemeris, we find that the TFA magnitudes are able to provide the best constraints on the transit parameters for HAT-P-67~b and thus we use these data for all of our following analyses. The corresponding log model evidence for each form of data is presented in Table \ref{tab:log-evidences}. 

We note that the data has been collected through two filters: the $I$ filter for the 2005 data and the $R_{C}$ filter for the 2008 data. Different photometric bandpasses yield different limb darkening solutions and instrumental systematics, and thus the common practice is to treat such data collected using the same instrument but different filters as two independent instruments. To test this hypothesis, we fit our data using \texttt{juliet} with uniform priors on the transit hyperparameters. We repeat the fit twice, once modeling the entire HATNet dataset with the same parameters and a second time stratifying the data by wavelength (i.e., defining separate median flux and jitter values for each filter). Indeed, we find that the stratification results in a higher log-evidence with $\Delta \ln \mathcal{Z} = 137.2$ ($34181.3$ versus $34044.1$; see Table \ref{tab:log-evidences}). In addition, since there are two photometric bandpasses, this results in an increase of only two parameters and does not significantly increase computation time for our final fits in \S\ref{sec:analysis}. 

\begin{deluxetable}{lcc}
\tablecaption{Log-evidences, a measure of how well the model fits the data, with better agreement corresponding to higher values, for fits to various combinations of photometric datasets.} \label{tab:log-evidences}
\tablehead{\colhead{Dataset} & \colhead{$\ln \mathcal{Z}$} & \colhead{Notes [\checkmark ?]}} 
\startdata 
    HATNet Raw &$30376.551$ &Single instrument\\
    HATNet EPD &$32070.551$ &Single instrument \\
    HATNet TFA &$34044.117$ &Single instrument\\
    HATNet TFA &$34181.261$ &Stratified by band [\checkmark]\\
    KeplerCam &$18534.661$ &Stratified by night\\
    KeplerCam &$18288.924$ &Single instrument [\checkmark]\\
    TESS & $442731.659$ &Single instrument [\checkmark]\\
    TESS & $442723.130$ &Stratified by sector
\enddata
\tablecomments{Configurations used in final modeling denoted by \checkmark}
\end{deluxetable}

\subsection{KeplerCam Follow-Up Photometry} \label{subsec:keplercam}

We made use of ground-based follow-up light curves from \citetalias{Zhou_2017} which were published alongside the discovery of HAT-P-67~b. These observations were carried out by KeplerCam on the Fred Lawrence Whipple Observatory (FLWO) 1.2m telescope and consist of 6 observations ranging from April 2011 to May 2013.\footnote{See \S2.1 of \citetalias{Zhou_2017} for a detailed description of these photometric observations.}

From the published magnitudes, we calculate the relative fluxes as well as flux uncertainties, and normalize each observation’s out-of-transit baseline flux to unity. Although the published data only contain magnitudes and no external parameters for detrending the data (e.g., airmass or full width at half maximum), the data quality is sufficiently high such that upon including \texttt{celerite}'s quasi-periodic Gaussian process (GP) \citep{ForemanMackey_2017} in our fits, we are largely able to match the precision reached by \citetalias{Zhou_2017}. 

Similar to the HATNet data, these data were collected in two bandpasses. The first five observations were obtained using the Sloan-\textit{i} filter, whereas the last observation used the Sloan-\textit{z} filter. Additionally, the six follow-up observations were obtained with slightly different exposure times. Though treating each night of data as collected by a different instrument nominally results in better log-evidence (see Table \ref{tab:log-evidences}), there was no observed improvement in the uncertainty of parameters and thus does not justify the doubling of the number of parameters (from $9$ to $19$) and the corresponding increase in computation time. 

\subsection{TESS Photometry} \label{subsec:tess}

The Transiting Exoplanet Survey Satellite \citep[TESS; ][]{10.1117/1.JATIS.1.1.014003} is an all-sky transit survey designed to observe planets transiting the nearest and brightest stars. TESS observed HAT-P-67~b in its 2-minute cadence mode in eight sectors (Sectors 24, 26, 51, 52, 53, 78, 79, and 80) between April 2020 and July 2024, all after the initial discovery in 2017. In total, 35 full and three partial transits were observed. 

We downloaded these data from the Mikulski Archive for Space Telescopes (MAST\footnote{\url{https://mast.stsci.edu/portal/Mashup/Clients/Mast/Portal.html}}) in the form of Pre-search Data Conditioning Simple Aperture Photometry (PDCSAP) light curves. This data has been corrected for instrumental systematics as well as photometric dilution by nearby sources, and a visual inspection of all sectors finds that no additional data removal is warranted. Additionally, we find that fitting all sectors of TESS with joint parameters for instrumental jitter and median flux results in the highest log-evidence as well as eight fewer parameters than stratifying by sector (see Table \ref{tab:log-evidences}). 

The high photometric precision of TESS, when combined with the datasets mentioned above, allows us to place more stringent constraints on the transit parameters. In addition, one important advantage offered by TESS is its continuous, high-precision out-of-transit monitoring which allows us to characterize stellar activity. We identify clear modulation in the light curve on the order of a fraction to a few parts per thousand which allows us to estimate the stellar rotation period of the star (see \S\ref{subsec:stellar}). This proves to be especially useful for this system, which has high scatter in the RVs due to the active, fast-rotating host. Whereas HATNet (see \S\ref{subsec:hatnet}) does provide out-of-transit monitoring, its precision of a few milli-magnitudes is not high enough to characterize this activity. On the other hand, ground-based follow-up (\S\ref{subsec:keplercam}) provides higher precision light curves but only for in-transit portions. Thus, besides offering improved uncertainties on planetary parameters as compared to previous data, TESS also offers a new perspective, allowing us to measure the stellar rotation period of the host star and gain a more robust understanding of the system. 

\section{Analysis} \label{sec:analysis}

\subsection{Stellar Properties} \label{subsec:stellar-info}

Knowledge of the host star's fundamental properties is crucial to transforming our differential measurements into physical quantities. The release of the Gaia DR3 parallax for the system also motivates a re-analysis of the host star. We inferred the fundamental and photospheric stellar parameters of
HAT-P-67 A using the \texttt{isochrones} \citep{mor15} package, 
executing a
simultaneous Bayesian fit of the Modules for Experiments in Stellar
Evolution \citep[MESA;][]{pax11,pax13,pax18,pax19,jer23} Isochrones \&
Stellar Tracks \citep[MIST;][]{dot16,cho16} isochrone grid to a curated
collection of data for the star with the nested sampler \texttt{MultiNest} \citep{fer08,fer09,fer19}. We fit the MIST grid to
\begin{enumerate}
\item
Galaxy Evolution Explorer (GALEX) GUVcat\_AIS FUV photometry including
in quadrature its zero-point uncertainty of 0.02 mag \citep{bia17};
\item
Gaia DR2 $G$ photometry including its zero-point uncertainty of 0.0018 mag 
\citep{gai16,gai18,are18,bus18,eva18,rie18};
\item
Tycho-2 $B_{T}$ and $V_{T}$ photometry including in quadrature their
zero-point uncertainties (0.078, 0.058) mag \citep{hog00,mar05};
\item
Two-micron All-sky Survey (2MASS) $JHK_{s}$ photometry. The 2MASS Point Source Catalog \texttt{?\_msigcom} photometric uncertainties we use, presented in \citet{skr06}, already includes zero-point uncertainties;
\item
Wide-field Infrared Survey Explorer (WISE) CatWISE2020 $W1$ and $W2$ photometry
including in quadrature their zero-point uncertainties (0.032, 0.037)
mag\footnote{\url{https://wise2.ipac.caltech.edu/docs/release/allsky/expsup/sec4\_4h.html\#PhotometricZP}}
\citep{wri10,mai11,eis20,mar21};
\item
a zero point-corrected Gaia DR3 parallax
\citep{gai21,fab21,lin21a,lin21b,row21,tor21}; and
\item
an estimated extinction value based on a three-dimensional extinction
map \citep{lal22,ver22}.
\end{enumerate}
As priors we use
\begin{enumerate}
\item
a \citet{cha03} log-normal mass prior for $M_{\ast} < 1~M_{\odot}$
joined to a \citet{sal55} power-law prior for $M_{\ast} \geq 1~M_{\odot}$;
\item
a metallicity prior based on the Geneva-Copenhagen Survey
\citep[GCS;][]{cas11};
\item
a log-uniform age prior between 1 Gyr and 10 Gyr;
\item
a uniform extinction prior in the interval 0 mag $< A_{V} < 0.5$ mag; and
\item
a distance prior proportional to volume between the \citet{bai21}
geometric distance minus/plus five times its uncertainty.
\end{enumerate}
The results of these analyses are shown in Table \ref{tab:isochrone} and Figure \ref{fig:host_corner}.

\begin{deluxetable}{lC}
\tablecaption{Adopted stellar parameters.}\label{tab:isochrone} 
\tablewidth{0pt}
\tablehead{
\colhead{Property (Unit)} & \colhead{Value}}
\startdata
\multicolumn{2}{l}{\textbf{Observed Quantities}} \\
GALEX $FUV$ & 19.74\pm0.13  \\
Gaia DR2 $G$      & 9.975\pm0.002 \\
Tycho $B_T$        & 10.629\pm0.087  \\
Tycho $V_T$        & 10.158\pm0.067  \\
2MASS $J$         & 9.145\pm0.021 \\
2MASS $H$         & 8.961\pm0.019 \\
2MASS $Ks$        & 8.900\pm0.019 \\
CatWISE $W1$      & 8.862\pm0.033 \\
CatWISE $W2$      & 8.877\pm0.038 \\
Gaia DR3 parallax (mas) & 2.701\pm0.010  \\
\hline
\multicolumn{2}{l}{\textbf{Isochrone-inferred parameters}$^\mathrm{a}$} \\
Effective temperature $T_{\mathrm{eff}}$ (K) & 6640^{+140}_{-130}  \\
Surface gravity $\log(g)~\mathrm{(cm~s^{-2})}$ & 3.84\pm0.03\\
Metallicity [Fe/H] & 0.14^{+0.14}_{-0.12}\\ 
Mass $M_{\ast}$ $(M_{\odot})$& 1.73\pm0.10 \\
Radius $R_{\ast}$ $(R_{\odot})$ & 2.62\pm0.03 \\
Luminosity $L_{\ast}$ $(L_{\odot})$& 11.96_{-0.73}^{+0.84} \\
Age $\tau_{\mathrm{iso}}$ (Gyr) & 1.46_{-0.26}^{+0.29} \\
Distance (pc) & 370.1\pm1.3 \\
\enddata
\tablecomments{Reported uncertainties for observed parameters are total (see \S\ref{subsec:stellar-info}). Isochrone-inferred parameters report the median and $1\,\sigma$ standard deviation drawn from the sampled posterior distribution. a: This work.}
\end{deluxetable}

\subsection{Photometry} \label{subsec:phot}

Following \S\ref{subsec:hatnet}, \ref{subsec:keplercam}, and \ref{subsec:tess}, we reached the following conclusions: 

\begin{itemize}
    \item using all three sources of photometric data (HATNet, TESS, and KeplerCam) results in the best constraints and most robust fit;
    \item HATNet data should be treated as two separate instruments, with different jitter and relative flux offsets for each bandpass;
    \item KeplerCam data should be treated as a single instrument, encompassing all six follow-up light curves; and
    \item all eight sectors of TESS data should also be treated as a single instrument, with the same jitter and flux offset values.
\end{itemize}

Although the ephemeris of HAT-P-67~b is already well-constrained from \citetalias{Zhou_2017}, our analysis includes the addition of the later TESS data, and thus we want to ensure that the priors for our joint fits encompass the updated $P$ and $t_0$ values. We performed a fit in \texttt{juliet} following the above points, and obtain $P = 4.8101076 \pm 0.0000011$ days and $t_0 = 2459010.99177 \pm 0.00017$ which are consistent with those from \citetalias{Zhou_2017}. 

\subsubsection{Stellar Activity From TESS} \label{subsec:stellar}

\begin{figure*}[ht!]
    \includegraphics[width=\textwidth]{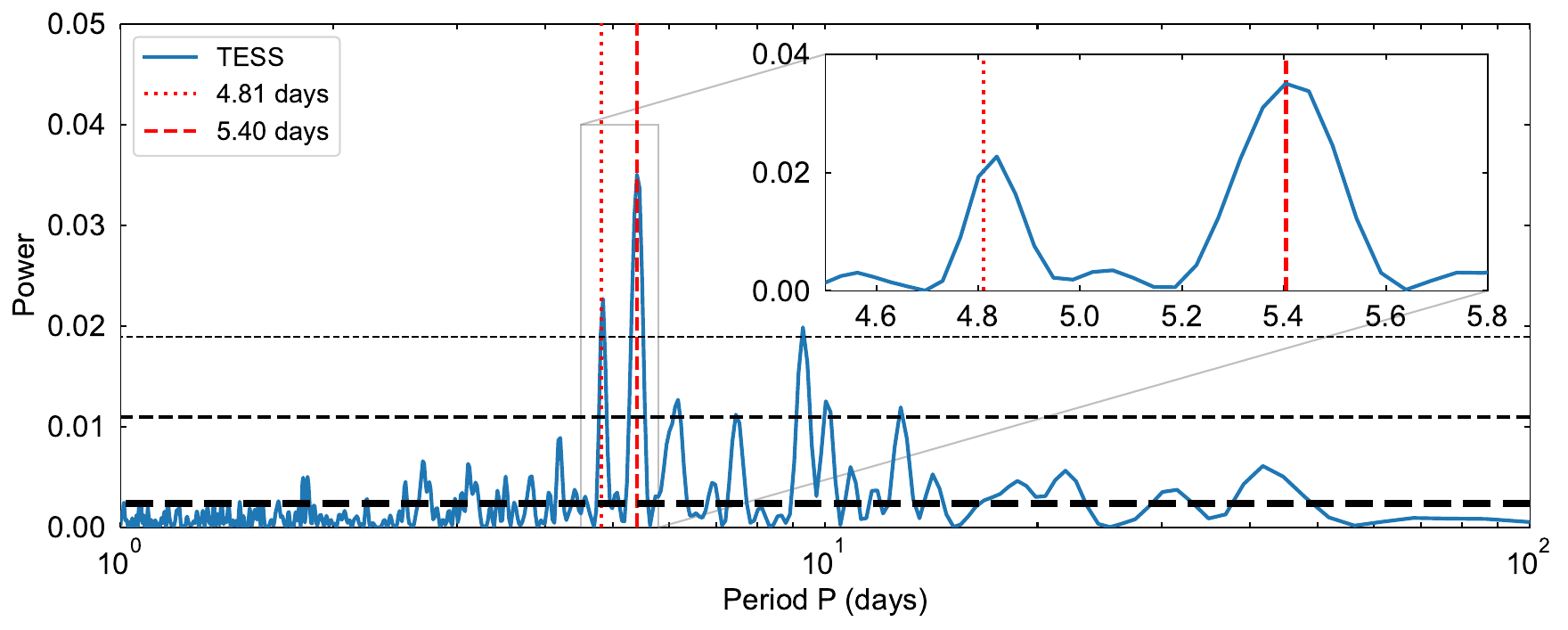}
    \caption{Periodogram of out-of-transit data from TESS (blue line). The Lomb-Scargle periodogram is designed to identify periodic signals within unevenly spaced data. Vertical red lines are drawn to indicate the period of the planet, $4.81$ days, and the suspected stellar rotation period of $5.40 \pm 0.09$ days. The horizontal lines represent $10\%$, $1\%$, and $0.1\%$ false-alarm levels calculated from 100 random draws. The inset shows the uncertainty on the stellar rotation peak and demonstrates that the data can conclusively resolve the stellar rotation and planetary signals. There is a signal at the planet's period even though its primary transits have been removed before calculating this periodogram.}
    \label{fig:tess_gls}
\end{figure*}

We aim to use TESS photometry to inform our model of the stellar activity in the RVs. From TESS, we observe quasi-periodic out-of-transit variability (see Figure \ref{fig:oot}). This variability is most obvious in the latter half of Sector 24 and in all of Sector 26. The amplitude of the variability is significantly less in Sectors 51 and 52. The activity returns to an elevated level after two years, during Sectors 78 through 80. With these clues, our original suspicion was that these arise from stellar activity due to starspots and/or faculae rotating in and out of view, and with said features having lifetimes on the order of the length of one TESS sector ($\sim 10 - 100$ days). To test this hypothesis, we plot the Lomb-Scargle periodogram \citep{Lomb_1976, Scargle_1982} of the TESS data, removing data within 0.05 in phase ($\sim 5.8$ hours) on both sides of each transit center to avoid capturing the transiting planetary signal. The periodogram is shown in Figure \ref{fig:tess_gls}, where we find that the variability has a period of $\sim 5$ days. 

\begin{figure}[ht!]
    \centering
    \includegraphics[width=\columnwidth]{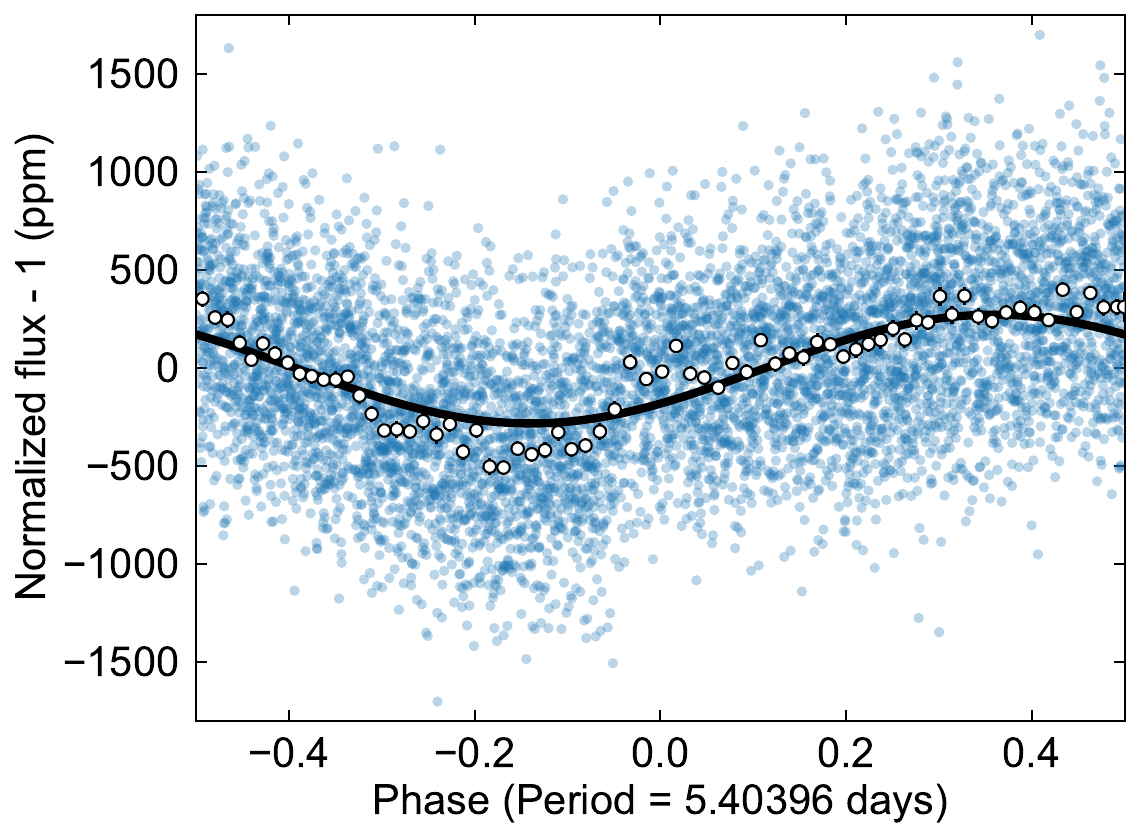}
    \caption{Out-of-transit data from TESS phased to the strongest peak of $5.40$ days. Blue points represent raw data binned by 10, and white circles represent bins of 1000. The best-fit sinusoidal model phased to the Lomb-Scargle peak is overplotted in black.}
    \label{fig:phased-oot}
\end{figure}

We observe the strongest peak (false-alarm probability $<0.1\%$) at $5.40 \pm 0.09$ days, where we obtain the uncertainty by fitting the peak with a Gaussian. Furthermore, upon phasing the photometry at this period, we measure an amplitude of $274 \pm 5$ parts per million (ppm). The phase-folded data is shown in Figure \ref{fig:phased-oot} and allows us to see the modulation. We believe this signal is due to stellar rotation, as upon conducting a periodogram analysis of Sector 26 data only, the sector which visually exhibits the strongest variability, we can recover the signal, with the strongest peak at $5.394$ days. Additionally, we find the strength of the signal decreases in Sectors 51 through 53 (see the full photometry in Figure \ref{fig:oot}).

Accounting for macroturbulence and instrumental broadening in addition to rotational broadening, \citetalias{Zhou_2017} finds ${v\sin i_{*}} = 30.9 \pm 2.0$ km/s and a macroturbulence of $9.22 \pm 0.5$ km/s for HAT-P-67 from HIRES spectra. Given the radius of HAT-P-67 ($R_{\ast} = 2.62 \pm 0.03~R_{\odot}$), a 5.4-day rotation period corresponds to $v=24$~km/s (consistent with the above value to $3\sigma$), assuming the stellar axis is approximately aligned with the plane of the sky. This assumption is justified as the planet has $i=85.0 ^{\circ}$, close to $90^{\circ}$, and \citetalias{Zhou_2017} measure through Doppler tomography that the system is spin-orbit aligned to within $12^{\circ}$, for a maximum deviation of $17^{\circ}$. Thus, our results suggest that more of the spectral line broadening is due to macroturbulence than found by \citetalias{Zhou_2017}. In other words, their macroturbulence estimate is low, leading to a $v\sin i_{*}$ measurement that is higher than ours by $3 \sigma$. HAT-P-67's rotation period is as expected for a star of its mass that is evolving off the main sequence \citep{van_Saders_2013}.

\subsubsection{Variability at Planetary Period} \label{subsubsec:phase_curve}

We note that in Figure \ref{fig:tess_gls}, despite having removed the primary transits from the data before constructing the periodogram, there is an additional, weaker peak at the period of the planet, which demonstrates that while the stellar rotation signal dominates the out-of-transit variability, there may be additional signs of variation such as a phase curve induced by HAT-P-67~b or secondary eclipses within the data. To test this, we phased the entire TESS light curve (including in-transit as well as out-of-transit) to the planetary period. We observe moderate phase curve modulations, but do not find clear evidence for a secondary. Reconstructing the periodogram in Figure \ref{fig:tess_gls} while excluding data within phase 0.05 of a hypothetical eclipse at phase 0.5 shows the same peak at 4.81 days, which is consistent with the lack of an observable secondary. We similarly did not find eclipse measurements of this target described in the literature. 

\begin{figure}
    \centering
    \includegraphics[width=1.0\linewidth]{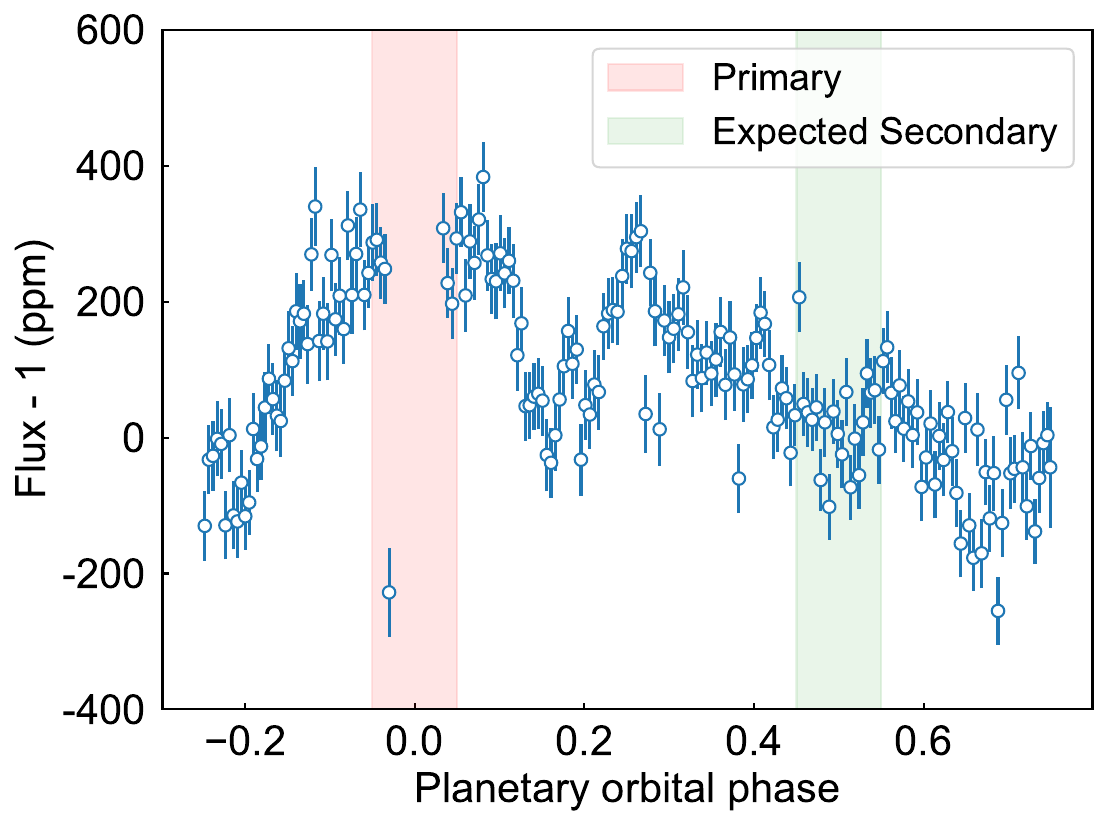}
    \caption{Light curve for HAT-P-67 from TESS, phased to the planetary orbital period. Each point here is the result of binning 400 2-minute cadence points in the TESS photometry. The light curve was first phase-folded then binned.}
    \label{fig:phase_curve}
\end{figure}

This is in agreement with theoretical calculations: given $R_{p}/R_{*} = 0.0823$, $T_{*} = 6417_{-143}^{+102}$ K, and $T_{p} = 1903 \pm 25$ K \citepalias{Zhou_2017}, the luminosity ratio between HAT-P-67~b and HAT-P-67 is $5.29 \times 10^{-5}$, for a secondary depth of $53$ ppm. While binning of 400 points allows us to reach a median errorbar of 52.6 ppm, there are too many variations of a few hundred ppm in the out-of-transit light curve that make it difficult to discern a secondary. Furthermore, the phase curve in Figure \ref{fig:phase_curve} exhibits steep ramps on both sides of primary transit, unlike traditional phase curves \citep[e.g.,][]{Daylan_2021}, which explains the power we observe at the planet period. We conclude that we may be observing signs of magnetic interactions between the star and planet, although a thorough analysis is outside the scope of this work (but see \S \ref{subsec:spmi}). 

\subsubsection{Transit Timing Variations} \label{subsubsec:ttvs}

\begin{figure*}[ht!]
    \includegraphics[width=\textwidth]{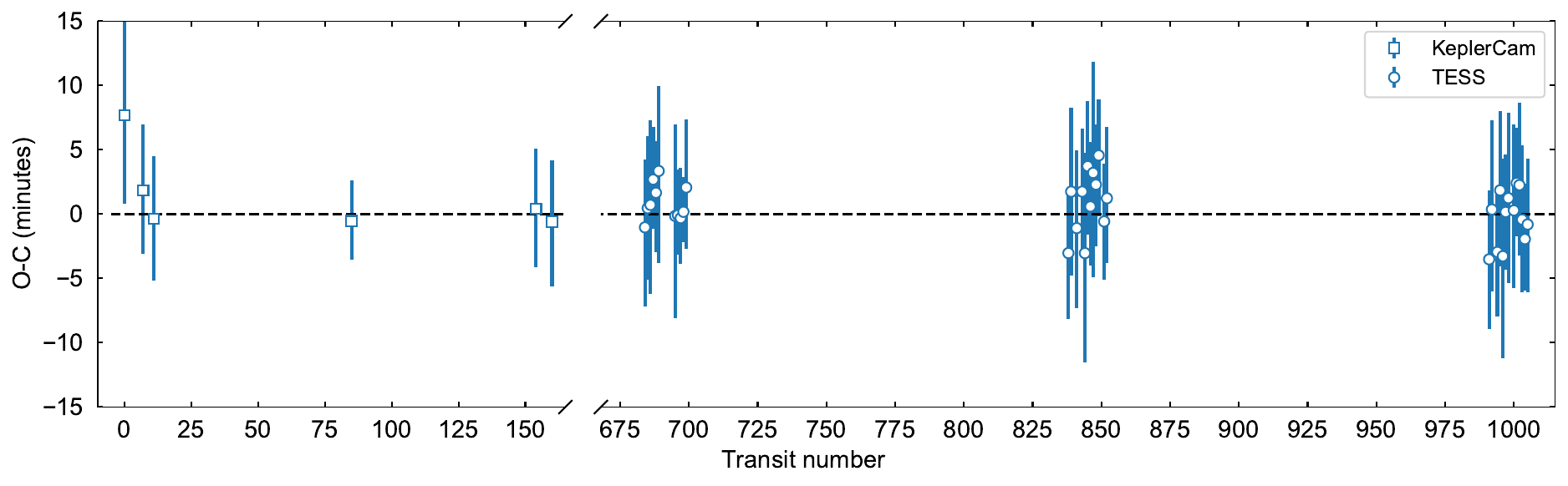}
    \caption{O-C plot of mid-transit times. No significant short- or long-term deviations are found. }
    \label{fig:ttvs}
\end{figure*}

As there may be interactions between the star and planet (\S \ref{subsec:spmi}), we search for orbital decay for HAT-P-67~b. We combine all transits observed by KeplerCam and TESS and fit each with an independent mid-time of transit. We exclude data from HATNet as its time resolution is insufficient to detect transits over a single orbital period. We define Gaussian priors for $t_0$ for each transit observed in the data, with medians equal to the predicted transit times assuming a linear ephemeris and standard deviations of $0.1$ day. The observed minus calculated transit times are plotted in Figure \ref{fig:ttvs}. We do not detect any significant timing variations on either short or long timescales, but this does not place significant constraints on the multiplicity of this system. 

\subsection{NEID Radial Velocities} \label{subsec:rvs}

\subsubsection{Periodogram of RVs} \label{3.3.2}

\begin{figure}
    \centering
    \includegraphics[width=\linewidth]{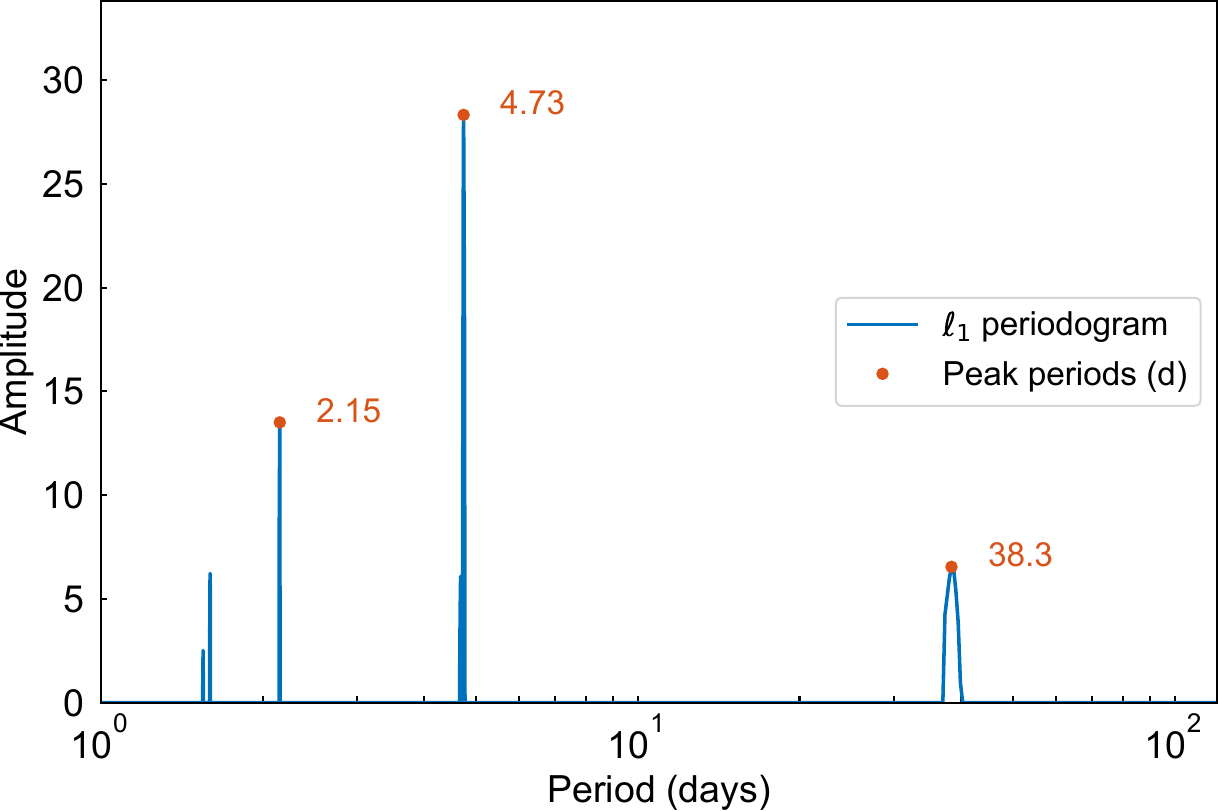}
    \caption{$l1$ periodogram for HAT-P-67 radial velocities. Labeled in orange are the periods of the peaks with Bayes factors $\geq 2.0$. The Bayes factors of the signals, in order of increasing period, are 6.7, 10.3, and 2.0 respectively. The p-values of the peaks are $2.85 \times 10^{-1}$, $2.55 \times 10^{-2}$, and $1.48 \times 10^{-4}$ respectively.}
    \label{fig:l1}
\end{figure}

To confirm the presence of the 4.8-day signal and to search for additional signals, we computed the $l1$ periodogram \citep{Hara_2016}, which is an implementation of the Least Angle Regression (LARS) algorithm \citep{efron_2004} and designed to search for planets using RVs. The $l1$ periodogram is more robust against aliasing than the Lomb-Scargle periodogram, thus resulting in fewer peaks, and aims to represent RV data as a sum of a few sinusoidal signals. The periodogram is plotted in Figure \ref{fig:l1}, with the periods of peaks with Bayes factors of $\geq 2.0$ labeled. The signal of the planet, $\approx 4.7$ days, is clearly represented. A signal of roughly half the planet period is present, although it is slightly less than half the planet period and is not significant (p-value of $0.285$). These suggest that it is either an alias of the planetary signal or a spurious signal. Additionally, there is a peak around $P=38$ days, but it is not immediately obvious whether this is a true frequency or an alias. 

\subsubsection{Two-Planet Fit} \label{additional}

Another method commonly used to search for additional signals in radial velocities is to perform a fit to the data including an extra planet, while allowing the period of the hypothetical planet to take on values in a broad, uniform range \citep[e.g.,][]{Espinoza2019_toi141}. To do this, we used the posteriors for $P$ and $t_0$ from our photometric fit as the priors for our first planet, effectively constraining the ephemeris of the known planet, and included a second planet with a free period. Specifically, we defined the period with a uniform prior between $1.0$ and $60$ days, as the largest continuous time span of our radial velocities is from February 2023 to July 2023 ($\sim 120$ days). We reason that it is justified to search for signals with periods half this length, but not more. We also define the $t_0$ prior for the second planet to be uniform, starting from the first data point and ending $60$ days later. Both ``planets'' were assumed to have circular orbits. 

Our results return $P = 5.048 \pm 0.660$ days for the second planet, with the posterior distribution exhibiting a single large peak at $P = 5.05$ days and isolated, weaker peaks at $P = 5.12$ and $13.0$ days. It should be noted that this two-planet fit performs nominally better than the single, 4.8-day planet fit, with a log-evidence that is higher by $9.8$ ($-498.193$ vs $-507.982$), despite having five more parameters. Statistically, this is ``strong'' evidence that there is an additional signal in the system, with the two-planet fit being $1.8 \times 10^4$ times more likely. 

However, it is physically improbable for HAT-P-67 to harbor two planets with periods of $P = 4.81$ and $P = 5.05$ days, as two such planets would be within each others' Hill radii. Instead, we note the close proximity of this second period with the period of out-of-transit variability obtained by TESS ($P = 5.40 \pm0.09$ days) and attribute these to originate from a single physical process. Interestingly, upon repeating the same fit on the first 66 of 75 RV points \citep[analogous to a cross-validation test; e.g.,][]{Blunt_2023}, the posterior distribution of $P$ exhibits a cluster of peaks around $P = 5.05$ days but also a larger cluster around $P = 5.6$ days. We interpret this as an indication that the period of the second signal in the data is not well-constrained by the RVs, but its range of values encompasses the signal seen in the photometry. Finally, upon repeating the two-planet fit and disallowing Hill sphere crossings, we obtain periods of $1.82^{+0.87}_{-0.42}$ and $12.95^{+0.40}_{-0.09}$ days. However, the log-evidences of these fits are significantly lower than for a single-planet fit, suggesting there is no evidence for planets at these periods. Combining the conclusions independently deduced from the photometry and RVs, we believe that it is justified to use a physically-motivated GP \citep{Nicholson_2022} when fitting the RVs in order to account for contamination from stellar activity \citep[e.g., ][]{Espinoza2019_toi141, Espinoza_2022}.

\subsubsection{Stellar Activity Indicators} \label{subsec:activity}

\begin{figure*}[ht!]
    \includegraphics[width=\textwidth]{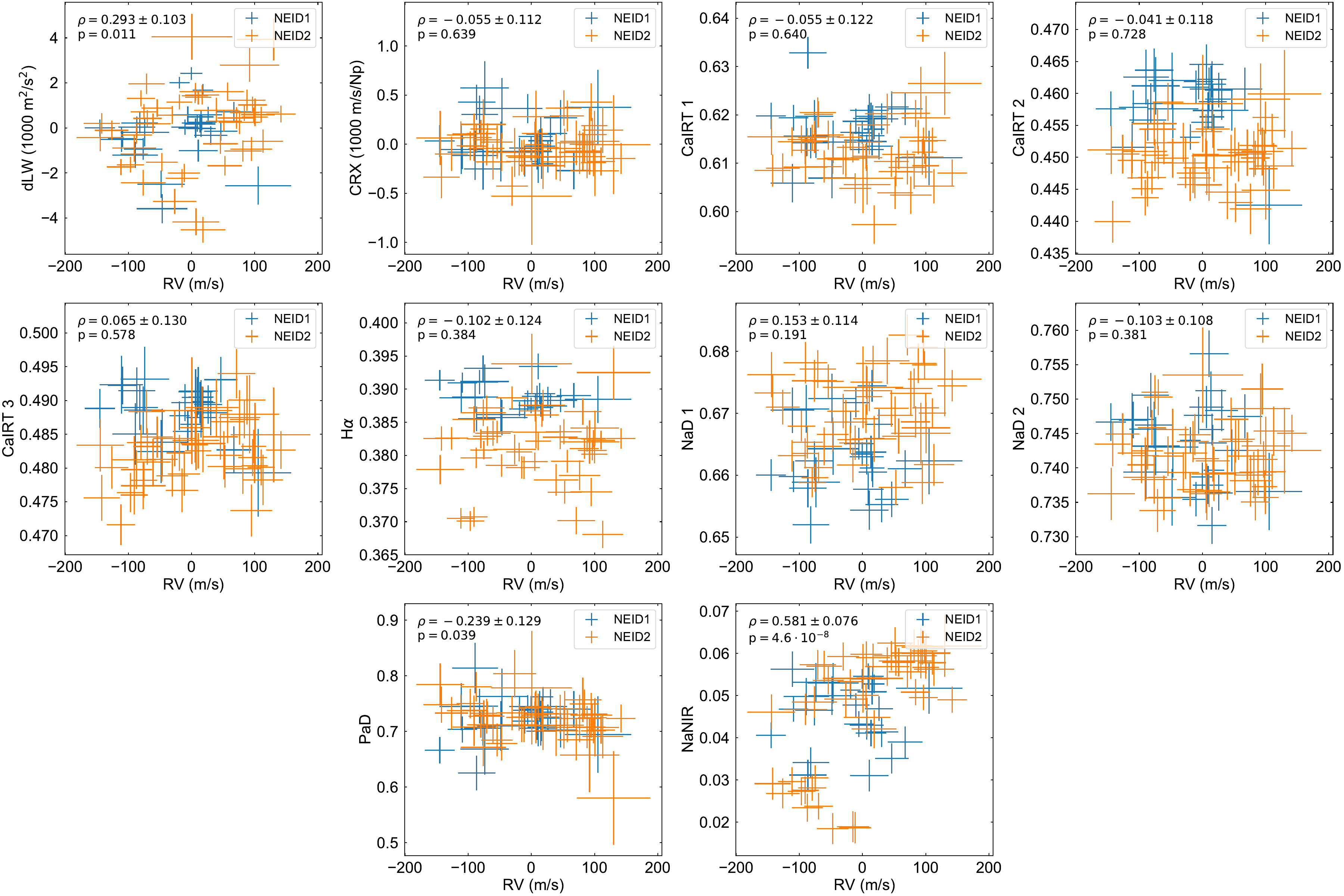}
    \caption{Correlations between the radial velocities and stellar activity indicators dLW, CRX, $\mathrm{Ca IRT}$, $\mathrm{H} \alpha$, $\mathrm{NaD}$,  $\mathrm{PaD}$, and $\mathrm{NaNIR}$ \citep[defined in][]{Zechmeister_2018}. NEID1 data is plotted in blue and NEID2 data in orange. The data from NEID1 and NEID2 are normalized to their respective zero-points. We calculate the uncertainties of the Spearman correlation coefficients using bootstrap resampling. The differential line width, Paschen delta, and sodium near-infrared indicators have statistically significant ($\mathrm{p}<0.05$) monotonic correlations with the radial velocities, with $\rho=0.293 \pm 0.101$, $-0.239 \pm 0.129$ and $0.581 \pm 0.076$ as obtained from a Spearman rank correlation test. A visual trend is not always immediately apparent due to the rank correlation test ignoring the actual values of the data.}
    \label{fig:correlations}
\end{figure*}

A total of 10 stellar activity indicators are calculated by the SERVAL pipeline \citep{Zechmeister_2018}. These are the differential line width (dLW), chromatic RV index (CRX), calcium infrared triplet ($\mathrm{CaIRT}$), $\mathrm{H} \alpha$, sodium D and near-infrared doublets, and Paschen delta indicators \citep[defined in][]{Zechmeister_2018}. We plot these activity indicators against the radial velocities in Figure \ref{fig:correlations}. We calculate the Spearman rank correlation coefficient $\rho$ between each indicator and the RVs, which tests for monotonic relationships between the two variables. We find that dLW yields $\rho = 0.293$, corresponding to $p=0.011$, demonstrating that the changing line width has a significant, positive correlation with the measured radial velocities and likely creates the observed scatter. It is worth noting that $\rho$ ignores the exact values of the data and only depends on the ranks of each variable, explaining why it is difficult to discern a trend from the dLW-RV panel in Figure \ref{fig:correlations}. We also find statistically significant correlations with the Paschen delta and sodium near-infrared doublet indicators.

\subsection{NEID + HIRES Radial Velocities}

Before we jointly fit the radial velocities with photometry, we conducted a few more analyses using all available radial velocity data to determine which parameters to fit for in our joint models. In particular, we investigate long-term linear trends and eccentricity.

\subsubsection{Linear Models}

Due to the presence of a co-moving, wide-binary companion HAT-P-67~B at 3400 AU \citep{10.1093/mnras/stz2673, gully_2023}, we test the possibility that there is a long-term linear trend in the radial velocity data. We fit using \texttt{juliet} a linear model to the radial velocity data with time as the regressor. However, the inclusion of such a model does not yield an increase in model log-evidence, nor are there significant non-zero linear terms (the posterior values for slope are consistent with zero). We conclude that HAT-P-67~B is too far away to cause a significant signal in the RVs within a period of $\sim 10$ years and do not use a linear model elsewhere in our modeling. 

\subsubsection{Eccentricity}

We also test the possibility that HAT-P-67~b may have an eccentric orbit by placing wide uniform priors on $e$ and $\omega$. However, the posterior distributions are skewed towards zero eccentricity, suggesting that the data do not provide constraints on $e$. \citetalias{Zhou_2017} reached similar conclusions based on their Doppler tomography and radial velocity results, and although we have additional data, we do not expect them to place constraints on $e$ due to our sampling scheme; eccentricity measurements are most constraining when sampling throughout the planet orbital phase. Nonetheless, due to the short orbital period of the planet, as well as the tidal circularization timescale of $\sim1$~Gyr \citep{Dobbs-Dixon_2004}, we do not believe it to have a significant eccentricity. Including the additional $e$ and $\omega$ parameters also does not increase the log-evidences of the fits, and thus for the purposes of our data modeling we fix the orbit to be circular in our final fits. 

\subsection{Global Analysis} \label{subsec:global}

\subsubsection{No Gaussian Process} \label{subsubsec:nogp}

\begin{figure*}[ht!]
    \includegraphics[width=\textwidth]{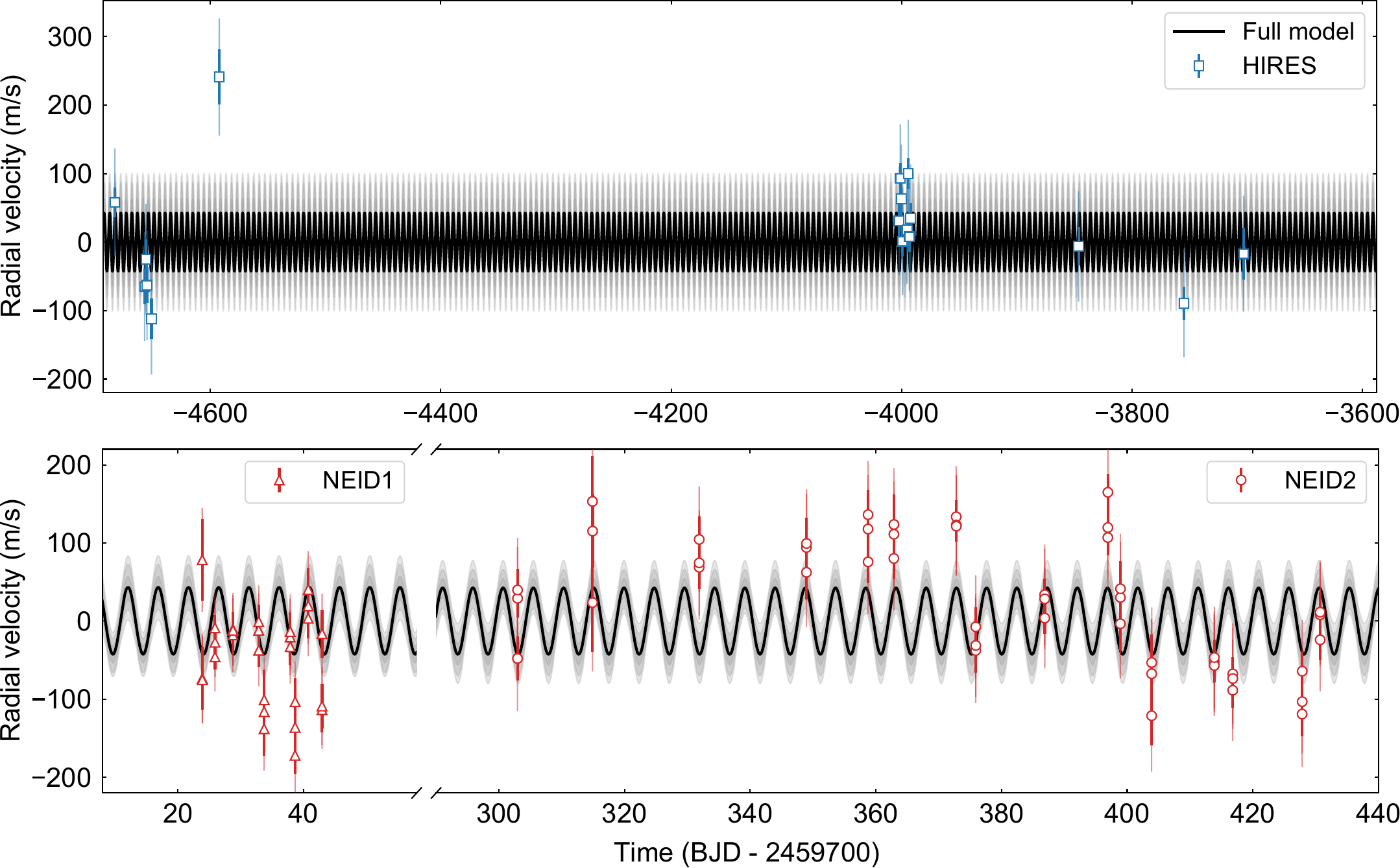}
    \caption{Radial velocity data points with overplotted model containing one Keplerian. The gray bands represent 68\%, 95\%, and 99\% confidence intervals. The data uncertainties are drawn as bold errorbars, and the thin errorbars represent the jitter added in quadrature to the data uncertainties. This model does not include a Gaussian process, resulting in large jitter values.}
    \label{fig:rvs-unphased}
\end{figure*}

\begin{figure*}[ht!]
    \includegraphics[width=\textwidth]{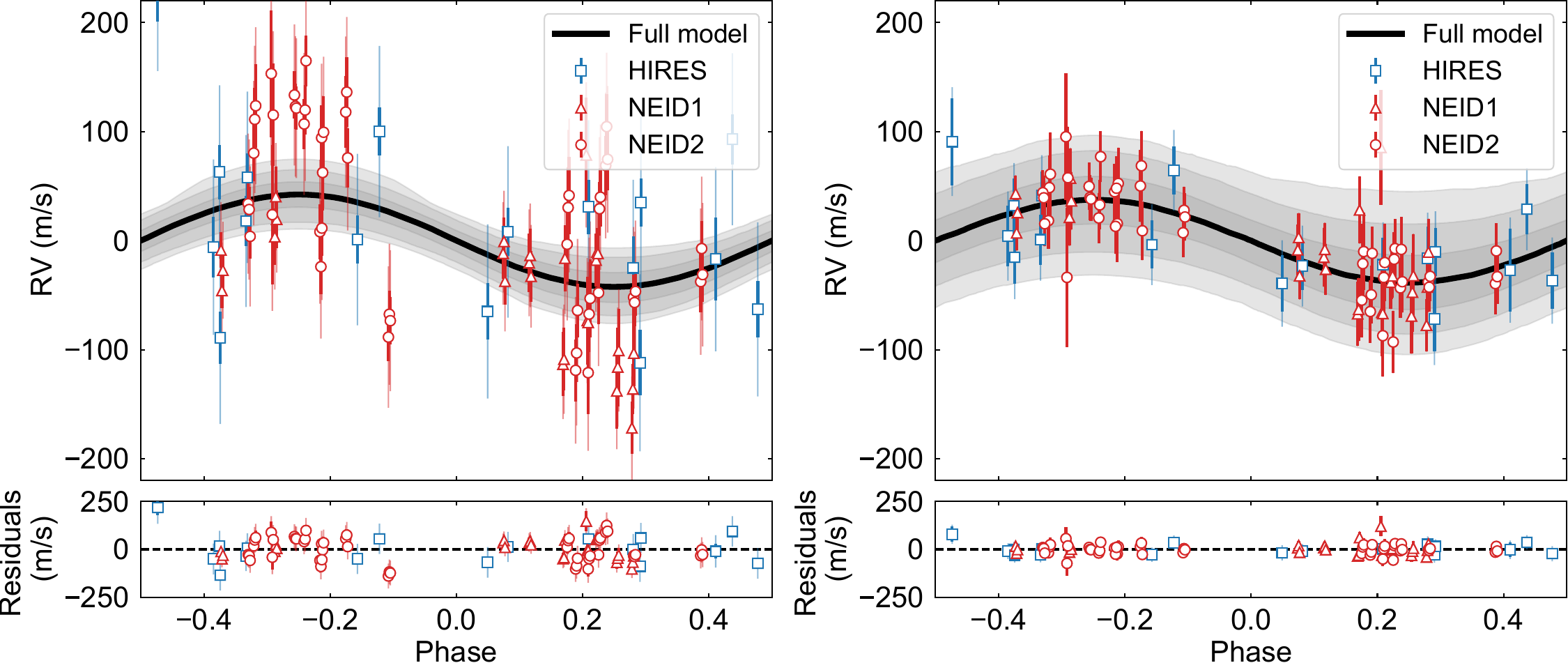}
    \caption{Phased RVs and best-fit model, without a Gaussian process (left) and with a quasi-periodic Gaussian process (right). The data points on the right are detrended by the Gaussian process component of the model, which effectively improves agreement between the data and the model.}
    \label{fig:rvs-phased}
\end{figure*}

Having gained an understanding of the photometric and RV data separately, we are now in a position to conduct joint fitting of this system including all datasets. Joint fitting gives self-consistent measurements of the system parameters, allowing us to derive HAT-P-67~b's density. We first conduct a joint fit which does not use a Gaussian process, and instead only includes planetary parameters for a single planet and instrumental nuisance parameters. Since we have demonstrated that the host star exhibits activity, we do not expect this fit to be the best fit to the data, and will also likely yield underestimated uncertainties. Nonetheless, it serves as a good baseline to which we can compare other models. We carry out our sampling using the \texttt{dynesty} \citep{Speagle2020} dynamic nested sampler to increase sampling efficiency. The priors and posteriors for each applicable parameter are listed in Table \ref{tab:planet-params}. The best-fit RV model is shown in Figure \ref{fig:rvs-unphased} and on the left panel in Figure \ref{fig:rvs-phased}.

\subsubsection{Quasi-Periodic Gaussian Process} \label{subsubsec:yesgp}

We next add \texttt{celerite}'s quasi-periodic Gaussian process, which consists of both quasi-periodic and exponential components. This is physically motivated by our previous analyses which show that the host star likely rotates at a period similar to the planet's orbital period. The priors for this fit are displayed in Table \ref{tab:planet-params}. To summarize, we take the best-fit values from our photometry-only fit (see \S\ref{subsec:phot}) for $P$ and $t_0$, and expand the standard deviation by a factor of $100$ to ensure we capture the true values. We allow $R_p/R_*$, $b$, $q_1$, and $q_2$ to vary uniformly between $0$ and $1$. Additionally, we implement a normal prior of $5.4 \pm 0.1$ days on the period of the quasi-periodic component of the GP, as a result of our analyses in \S\ref{subsec:phot} and \ref{subsec:rvs}. We find that this prior is able to capture the stellar rotation period while excluding the planetary period. We implement broad log-uniform priors on the remaining hyperparameters of the GP (amplitude, additive factor, and exponential length-scale). A description of the hyperparameters can be found in \citet{Espinoza2019}. 

\begin{figure*}[ht!]
    \includegraphics[width=\textwidth]{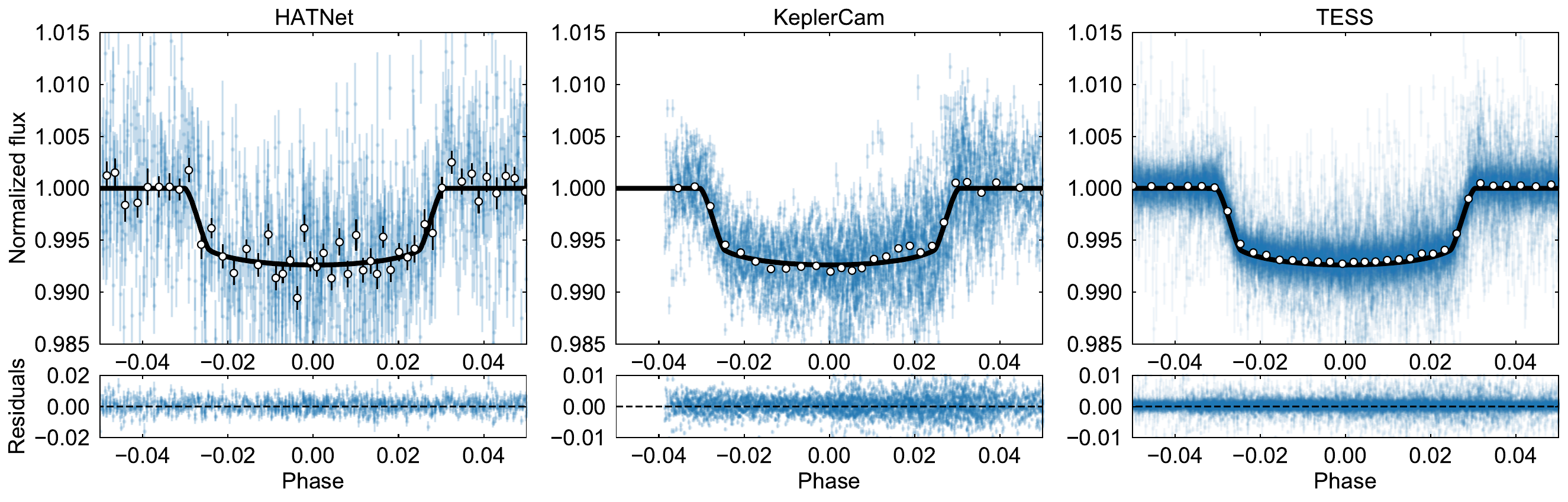}
    \caption{Phased raw (blue) and binned (white circles) photometry data points. Best-fit models (black lines) and residuals are also included.}
    \label{fig:phot-phased}
\end{figure*}

The unphased and phased radial velocities are shown in Figures \ref{fig:rvs-unphased} and \ref{fig:rvs-phased} respectively. The phased and fitted photometry is shown in Figure \ref{fig:phot-phased}. We find that the log-evidences are greatly increased upon using a quasi-periodic GP as compared to using no GP (747223 vs 730107, or $\Delta \ln \mathcal{Z} = 17116$). The posteriors are listed in Table \ref{tab:planet-params} and the corner plot is shown in Figure \ref{fig:corner}. 

\begin{deluxetable*}{lcccr}
    \tablecaption{Priors and posteriors for joint fit of HAT-P-67~b.} \label{tab:planet-params}
    \tablehead{\colhead{Parameter} & \colhead{Prior} & \colhead{QP GP Posterior} & \colhead{No GP Posterior} & \colhead{Description}}
    \startdata 
        \textbf{Planetary Parameters} \\
        $P$       &$\mathcal{N} (4.8101, 0.0001)$  & $4.81010690^{+0.00000070}_{-0.00000068} $ & $4.81010827^{+0.00000058}_{-0.00000059} $ & Orbital period (days)\\
        $t_{0}$      &$\mathcal{N} (2011.0, 0.1)$    & $2010.99174 ^{+0.00017}_{-0.00016}$  & $2010.99156 ^{+0.00015}_{-0.00014}$ & Transit center ($\mathrm{BJD}_{\mathrm{TDB}} - 2457000$)\\
        $R_{p}/R_{\ast}$       &$\mathcal{U} (0.0,1.0)$           & $0.08200^{+0.00025}_{-0.00026}$ & $0.08059^{+0.00028}_{-0.00025}$ & Planet-star radius ratio\\
        $b$       &$\mathcal{U} (0.0,1.0)$           & $ 0.460^{+0.017}_{-0.019}$  & $0.433^{+0.023}_{-0.024}$ & Impact parameter\\
        $a/R_{\ast}$  &$\mathcal{J} (1.0, 100.0)$        & $5.173^{+0.049}_{-0.046}$  & $5.273^{+0.059}_{-0.058}$ & Scaled semi-major axis\\
        $e$       &Fixed, $0.0$     & ---        & --- & Eccentricity\\
        $\omega$  &Fixed, $90.0$    & ---        & --- & Argument of periastron ($^{\circ}$) \\
        $K$       &$\mathcal{U} (0.0, 1000.0)$  & $38^{+12}_{-13}$ & $43 \pm 8$ & Semi-amplitude (m/s)\\
        $M_p$       & --- & $0.45 \pm 0.15$ & --- & Mass (M\textsubscript{J})\\
        $R_p$       & --- & $2.140 \pm 0.025$ & --- & Radius (R\textsubscript{J})\\
        $\rho_p$       & ---  & $0.061^{+0.020}_{-0.021}$ & --- & Density ($\mathrm{g} \mathrm{~cm}^{-3}$)\\
        \hline
        \textbf{Photometric Parameters} \\
        $D_{\mathrm{HATNet}}$    &$\mathcal{U} (0.0,1.0)$    &$0.752 ^{+0.028}_{-0.027}$    & $0.770^{+0.030}_{-0.028}$    &Dilution factor for HATNet\\
        $F_{\mathrm{TESS}}$    &$\mathcal{N} (0.0, 0.1)$    &$-90 \pm 170$    & $-89 \pm 3$    & Relative flux offset (ppm)\\
        $F_{\mathrm{KeplerCam}}$    &$\mathcal{N} (0.0, 0.1)$    &$150 \pm 340$    & $173^{+48}_{-50}$    & Relative flux offset (ppm)\\
        $F_{\mathrm{HATNet}}$    &$\mathcal{N} (0.0, 0.1)$    &$-90^{+770}_{-800}$     & $-13 \pm 65$     & Relative flux offset (ppm)\\
        $\sigma_{w\mathrm{,\,TESS}}$    &$\mathcal{J} (10^{-5}, 10^{5})$ &$0.014^{+1.231}_{-0.014}$ & $479.5^{+4.4}_{-4.5}$ & Jitter term for TESS (ppm)\\
        $\sigma_{w\mathrm{,\,KeplerCam}}$    &$\mathcal{J} (10^{-5}, 10^{5})$ &$2746^{+30}_{-32}$ & $3019 \pm 36$ & Jitter term for KeplerCam (ppm)\\
        $\sigma_{w\mathrm{,\,HATNet}}$    &$\mathcal{J} (10^{-5}, 10^{5})$ &$4450^{+33}_{-32}$ & $4466 \pm 34$ & Jitter term for HATNet (ppm)\\
        $q_1$          &$\mathcal{U} (0.0,1.0)$           & $0.173 ^{+0.026}_{-0.024} $ & $0.159^{+0.030}_{-0.027} $ & First limb-darkening coefficient\\
        $q_2$          &$\mathcal{U} (0.0,1.0)$           & $0.29^{+0.08}_{-0.07}$ & $0.38^{+0.10}_{-0.09}$ & Second limb-darkening coefficient\\
        $B_{\mathrm{TESS}}$    &$\mathcal{J} (10^{-6}, 10^{6})$  & $1.04^{+0.06}_{-0.03}$ & --- &Amplitude of GP ($10^{-6}$ ppm)\\
        $B_{\mathrm{KeplerCam}}$    &$\mathcal{J} (10^{-6}, 10^{6})$  & $1.92^{+0.75}_{-0.47}$ & --- &Amplitude of GP ($10^{-6}$ ppm)\\
        $B_{\mathrm{HATNet}}$    &$\mathcal{J} (10^{-6}, 10^{6})$  & $ 1.31^{+0.71}_{-0.24}$ & --- &Amplitude of GP ($10^{-6}$ ppm)\\
        $C_{\mathrm{TESS}}$    &$\mathcal{J} (10^{-4}, 10^{4})$  & $2.4^{+2.9}_{-1.7} $ & --- &Additive factor for GP amplitude\\
        $C_{\mathrm{KeplerCam}}$    &$\mathcal{J} (10^{-4}, 10^{4})$  & $0.42^{+216.37}_{-0.42}$ & --- &Additive factor for GP amplitude\\
        $C_{\mathrm{HATNet}}$    &$\mathcal{J} (10^{-4}, 10^{4})$  & $ 0.25^{+14.70}_{-0.24}$ & --- &Additive factor for GP amplitude\\
        $L_{\mathrm{TESS}}$    &$\mathcal{J} (10^{-3}, 10^{3})$  & $5.03^{+0.38}_{-0.29}$ & --- &Exponential length-scale\\
        $L_{\mathrm{KeplerCam}}$    &$\mathcal{J} (10^{-3}, 10^{3})$  & $0.075^{+0.038}_{-0.022} $ & --- &Exponential length-scale\\
        $L_{\mathrm{HATNet}}$    &$\mathcal{J} (10^{-3}, 10^{3})$  & $ 290^{+360}_{-170} $ & --- &Exponential length-scale\\
        $P_{\mathrm{rot,\,TESS}}$    &$\mathcal{N} (5.4, 0.1)$  & $5.39\pm 0.09 $ & --- &Period for QP component (days)\\
        $P_{\mathrm{rot,\,KeplerCam}}$    &$\mathcal{N} (5.4, 0.1)$  & $5.38\pm 0.09 $ & --- &Period for QP component (days)\\
        $P_{\mathrm{rot,\,HATNet}}$    &$\mathcal{N} (5.4, 0.1)$  & $ 5.35\pm 0.07 $ & --- &Period for QP component (days)\\
        \hline
        \textbf{RV Parameters}\\
        $\mu_{\mathrm{NEID1}}$ & $\mathcal{U} (-200,200)$ &$-25^{+31}_{-30}$ & $-28 \pm 10$ & RV zero-point for NEID1 (m/s)\\
        $\mu_{\mathrm{NEID2}}$ & $\mathcal{U} (-200, 200)$ &$24^{+19}_{-20}$ & $23 \pm 10$ & RV zero-point for NEID2 (m/s)\\
        $\mu_{\mathrm{HIRES}}$ & $\mathcal{U} (-200, 200)$ &$15^{+28}_{-26}$ & $15 \pm 19$ & RV zero-point for HIRES (m/s)\\
        $\sigma_{\mathrm{NEID1}}$ & $\mathcal{J}(1.0, 1000.0)$ &$4.6^{+7.9}_{-3.0}$ & $40.9^{+9.4}_{-7.9}$ & Jitter term for NEID1 (m/s)\\
        $\sigma_{\mathrm{NEID2}}$ & $\mathcal{J}(1.0, 1000.0)$  &$2.9^{+3.5}_{-1.5}$ & $60.9^{+8.5}_{-6.7}$ & Jitter term for NEID2 (m/s)\\
        $\sigma_{\mathrm{HIRES}}$ & $\mathcal{J}(1.0, 1000.0)$  &$30^{+32}_{-20}$ & $75^{+18}_{-14}$ & Jitter term for HIRES (m/s)\\
        $B_{\mathrm{NEID1+NEID2}}$    &$\mathcal{J} (1.0, 10^{6})$  & $4200 ^{+1900}_{-1200}$ & --- &Amplitude of GP (m/s)\\
        $B_{\mathrm{HIRES}}$    &$\mathcal{J} (1.0, 10^{6})$  & $5300 ^{+5600}_{-3600}$ & --- &Amplitude of GP (m/s)\\
        $C_{\mathrm{NEID1+NEID2}}$    &$\mathcal{J} (0.01,100)$  & $0.4^{+6.4}_{-0.3}$ & --- &Additive factor for GP amplitude\\
        $C_{\mathrm{HIRES}}$    &$\mathcal{J} (0.01,100)$  & $3.5^{+31.4}_{-3.4}$ & --- &Additive factor for GP amplitude\\
        $L_{\mathrm{NEID1+NEID2}}$    &$\mathcal{J} (1.0,100)$  & $4.1^{+7.7}_{-2.6}$ & --- &Exponential length-scale\\
        $L_{\mathrm{HIRES}}$    &$\mathcal{J} (1.0,100)$  & $10.6^{+25.2}_{-7.7}$ & --- &Exponential length-scale\\
        $P_{\mathrm{rot,\,NEID1+NEID2}}$    &$\mathcal{N} (5.4, 0.1)$  & $5.39 \pm 0.09$ & --- &Period for QP component (days) \\
        $P_{\mathrm{rot,\,HIRES}}$    &$\mathcal{N} (5.4, 0.1)$  & $5.39 \pm 0.09$ & --- &Period for QP component (days)
    \enddata
\end{deluxetable*}

\subsubsection{Uncertainty Comparison} \label{subsubsec:comparison}
The uncertainty on the RV semi-amplitude is increased from $19\%$ ($43 \pm 8$~m/s) to $33\%$ ($38^{+12}_{-13}$~m/s) upon including a Gaussian process; we believe this larger uncertainty more accurately reflects unknown noise sources in the data. In particular, upon inclusion of a GP, the jitter values have decreased significantly (from $40.9^{+9.4}_{-7.9}$~m/s to $4.6^{+7.9}_{-3.0}$~m/s for NEID1, and from $60.9^{+8.5}_{-6.7}$~m/s to $2.9^{+3.5}_{-1.5}$~m/s for NEID2). Together with the increase in uncertainty on $K$, we interpret this to indicate the uncertainties of the model have been correctly inflated to account for the excess discrepancies between the data and model, rather than inflating the data uncertainties. 

\begin{figure*}[ht!]
    \includegraphics[width=\textwidth]{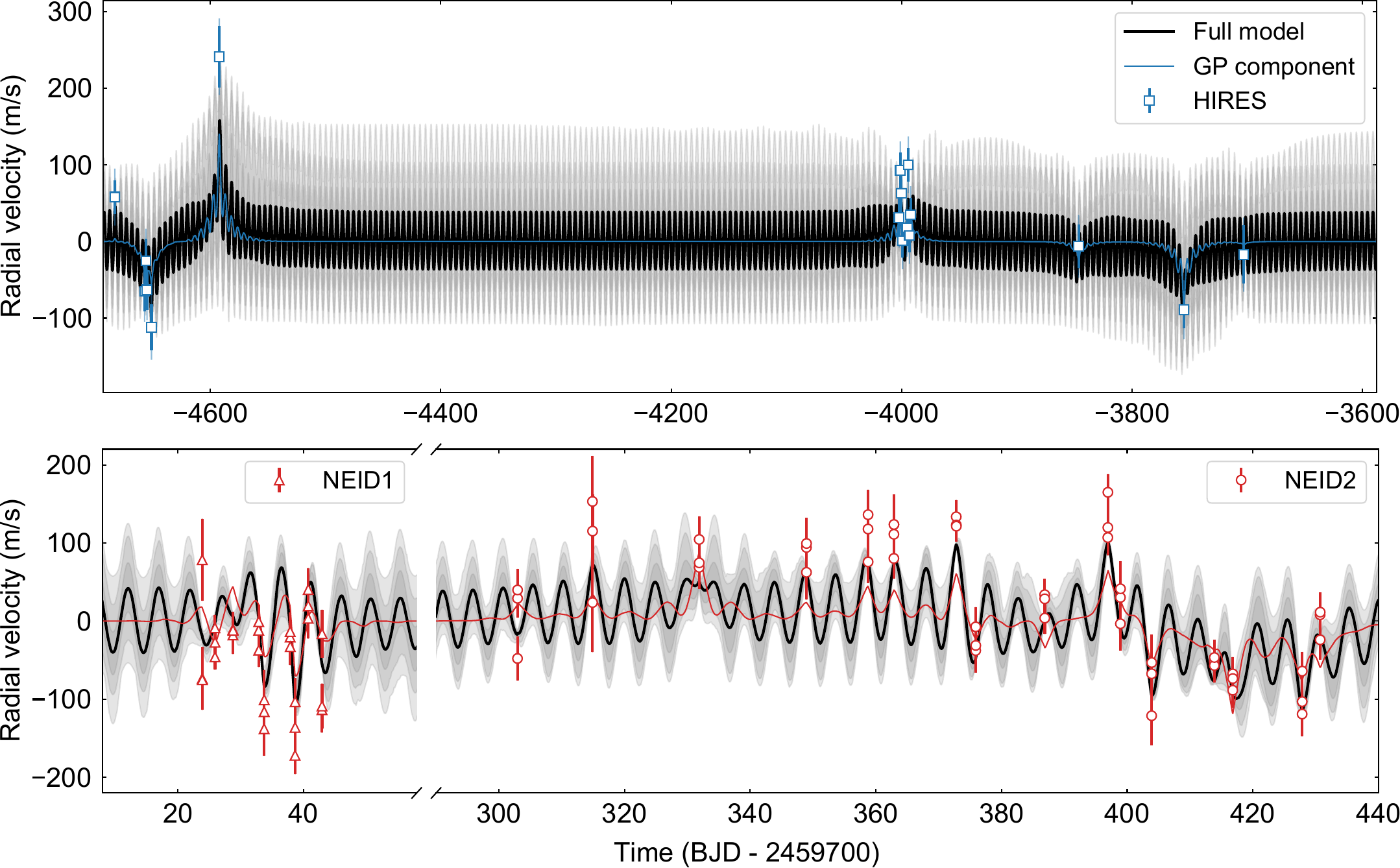}
    \caption{Same as Figure \ref{fig:rvs-unphased}, but for the full model which includes a by-instrument quasi-periodic Gaussian process component. The blue and red lines represent the GP component for the HIRES and NEID data respectively.}
    \label{fig:rvs-unphased-gp}
\end{figure*}

\section{Discussion}\label{sec:discussion}

\subsection{Implications for HAT-P-67~b}

Our revised mass measurement of $M_p = 0.45 \pm 0.15$~M\textsubscript{J} is consistent with the previous upper-limit of $M_p <0.59$~M\textsubscript{J} obtained by \citetalias{Zhou_2017}. When combined with our updated stellar mass calculation, this yields $\rho_p = 0.061^{+0.020}_{-0.021} \mathrm{~g} \mathrm{~cm}^{-3}$, also consistent with $\rho_p=0.052_{-0.028}^{+0.039} \mathrm{~g} \mathrm{~cm}^{-3}$ from \citetalias{Zhou_2017} and with a smaller uncertainty. The mass uncertainty of $33\%$ places the object in a region where the mass uncertainty and spectroscopic data quality have comparable contributions to uncertainty in inferred atmospheric properties from transmission spectroscopy, and lies in the regime between initial and detailed atmospheric characterization as suggested by \citet{Batalha2019}, which was the initial goal of the NEID observing strategy.

\begin{figure}
    \centering
    \includegraphics[width=\columnwidth]{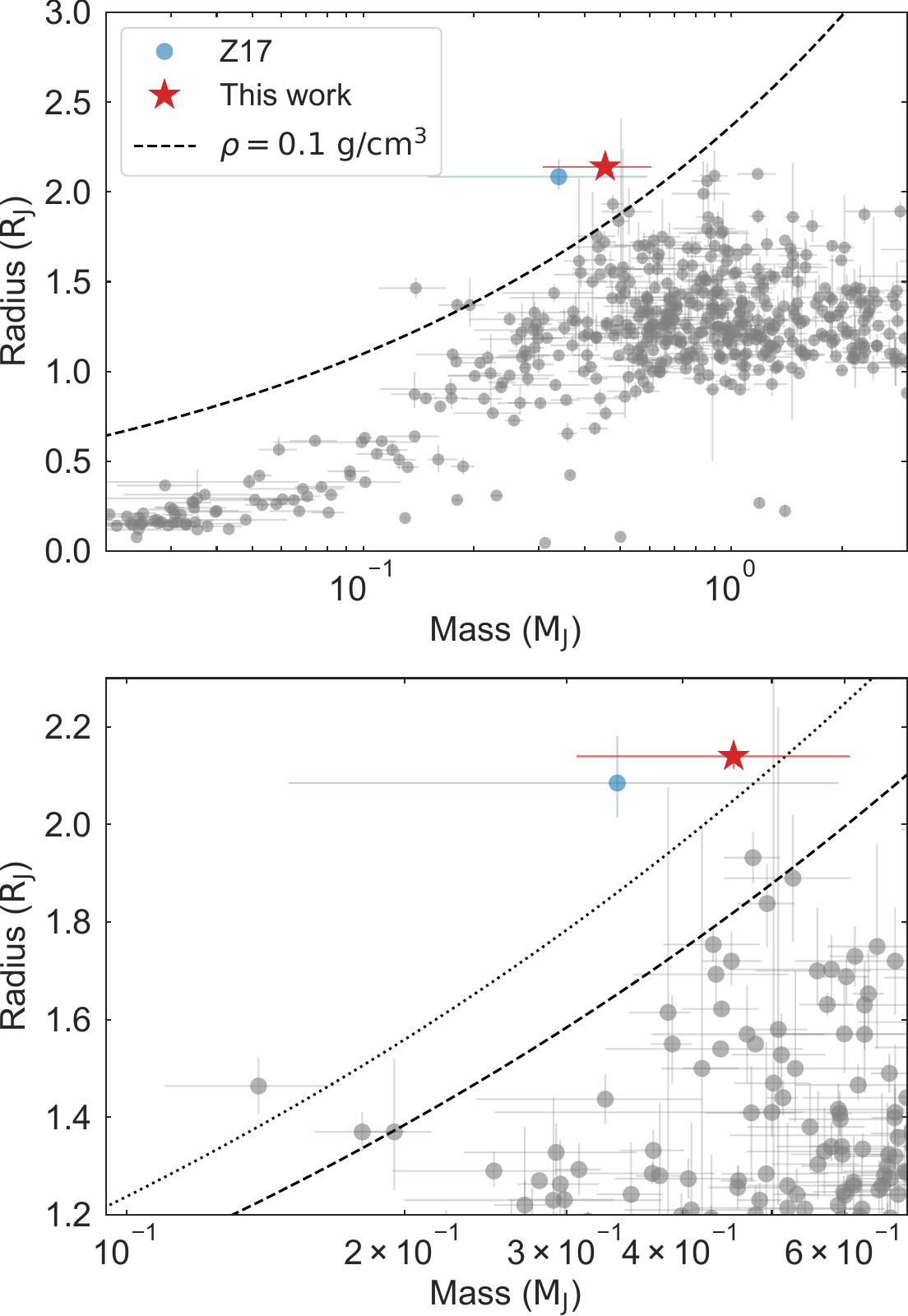}
    \caption{\textbf{Top:} Mass-radius plot of known transiting exoplanets with equilibrium temperatures greater than $1000$~K. The original location of HAT-P-67~b from \citetalias{Zhou_2017} is marked by a blue shaded circle and the updated location from this work by a red star. HAT-P-67~b is the largest known hot giant planet, and is one of only four planets to cross the $\rho = 0.1 \mathrm{~g} \mathrm{~cm}^{-3}$ boundary. \textbf{Bottom:} Zoomed-in version, with curves representing $\rho = 0.07 \mathrm{~g} \mathrm{~cm}^{-3}$ and $0.1 \mathrm{~g} \mathrm{~cm}^{-3}$. The only other planet with $\rho < 0.07 \mathrm{~g} \mathrm{~cm}^{-3}$ is WASP-193~b.}
    \label{fig:mprp}
\end{figure}

This revised density maintains HAT-P-67~b's position as one of the lowest-density planets known, placing it well outside the boundaries of the mass-radius distribution for known hot giants (see Figure \ref{fig:mprp}). It is the largest known planet, and its nominal density makes it the second least dense transiting planet with $T_{\mathrm{eq}} > 1000$~K. 

Models from \citet{Thorngren_2023} suggest that a hot Saturn in the current location of HAT-P-67~b in mass-radius space will lose the majority of its gaseous envelope within $\sim 0.05$ Gyr, after which it will retain only its core of $5-15~M_{\oplus}$ \citep{gully_2023}. 

\subsection{Coincidence of Stellar Rotation and Planetary Orbit} \label{subsec:spmi}

\par The proximity of the measured stellar rotation period of $5.4\pm0.1\,\mathrm{days}$ to the planet's orbital period of $4.8$ days could be coincidental, or due to secular tidal interactions between the star and planet \citep{gully_2023}. Alternatively, a more complicated interaction between the planetary magnetic field and the star could cause this coincidence of periods \citep[e.g., Star-Planet Magnetic Interaction, or SPMI;][]{Strugarek2018}. Correlation between stellar activity indicators and the planetary orbital period can be indicative of SPMI, but this correlation is not always present, as changes in the stellar magnetic field can produce periods of stellar rotation correlated activity and orbital period correlated activity \citep[e.g.,][]{Shkolnik2008}. The SPMI interpretation is disfavored by \citet{gully_2023} because they do not observe significant variability in the H$\alpha$ and Ca II HK lines. However, our data suggests there is excess periodogram power in the H$\alpha$ line and stellar photometry, both at the planetary period, which could possibly be an indication of planetary orbit correlated activity.

\subsection{Stellar Evolution} \label{subsec:evolution}

\begin{figure}[ht!]
    \centering
    \includegraphics[width=\columnwidth]{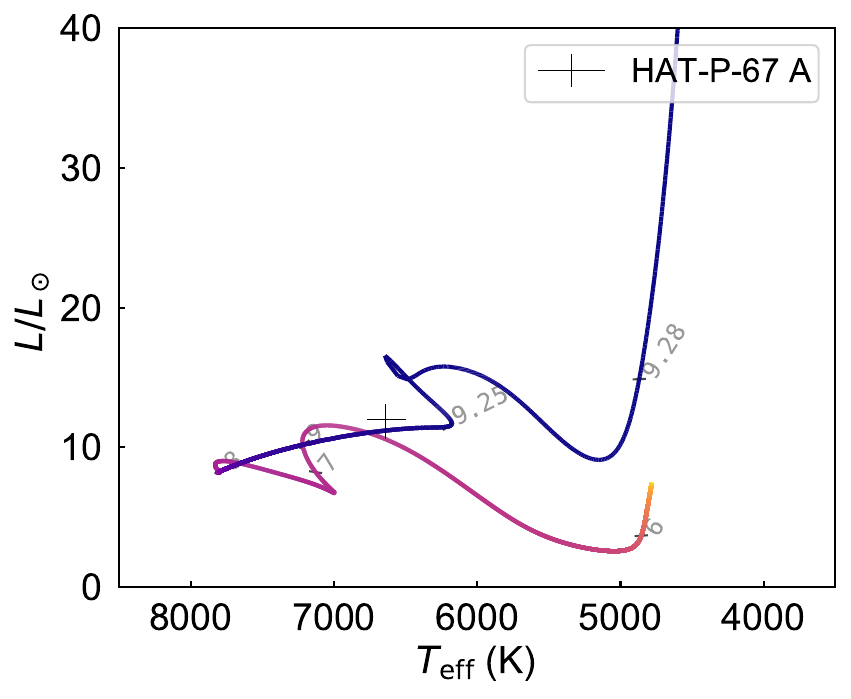}
    \caption{Evolution of a $1.73~M_{\odot}$ star from MESA in luminosity-$T_{\mathrm{eff}}$ space. The current parameters of HAT-P-67 A are plotted as the black cross, and the gray numbers represent the logarithm of the age of the star in years.}
    \label{fig:lteff}
\end{figure}

Due to HAT-P-67's large current radius and its expected further inflation as it continues to evolve off the main sequence, its radius will eventually increase past the orbit of HAT-P-67~b, thus engulfing the planet. To estimate the timescale of such an event, we use MESA 1-D stellar evolutionary models for a star at the mass of HAT-P-67 A ($1.73~M_{\odot}$). The MESA models are configured following a similar methodology as described in L. Wang \& K. Schlaufman (in prep). Using the median stellar parameters derived from our isochrone fit as inputs, we evolve the host star through its evolutionary phases up to the onset of the core helium flash. Additionally, the luminosity of the host star will dramatically increase in the near future, accelerating the planet's evaporation. The modeled luminosity over time is shown in Figure \ref{fig:lteff}. 

\subsection{Tidal Evolution} \label{subsec:tidal}

\begin{figure*}
    \centering
    \includegraphics[width=\linewidth]{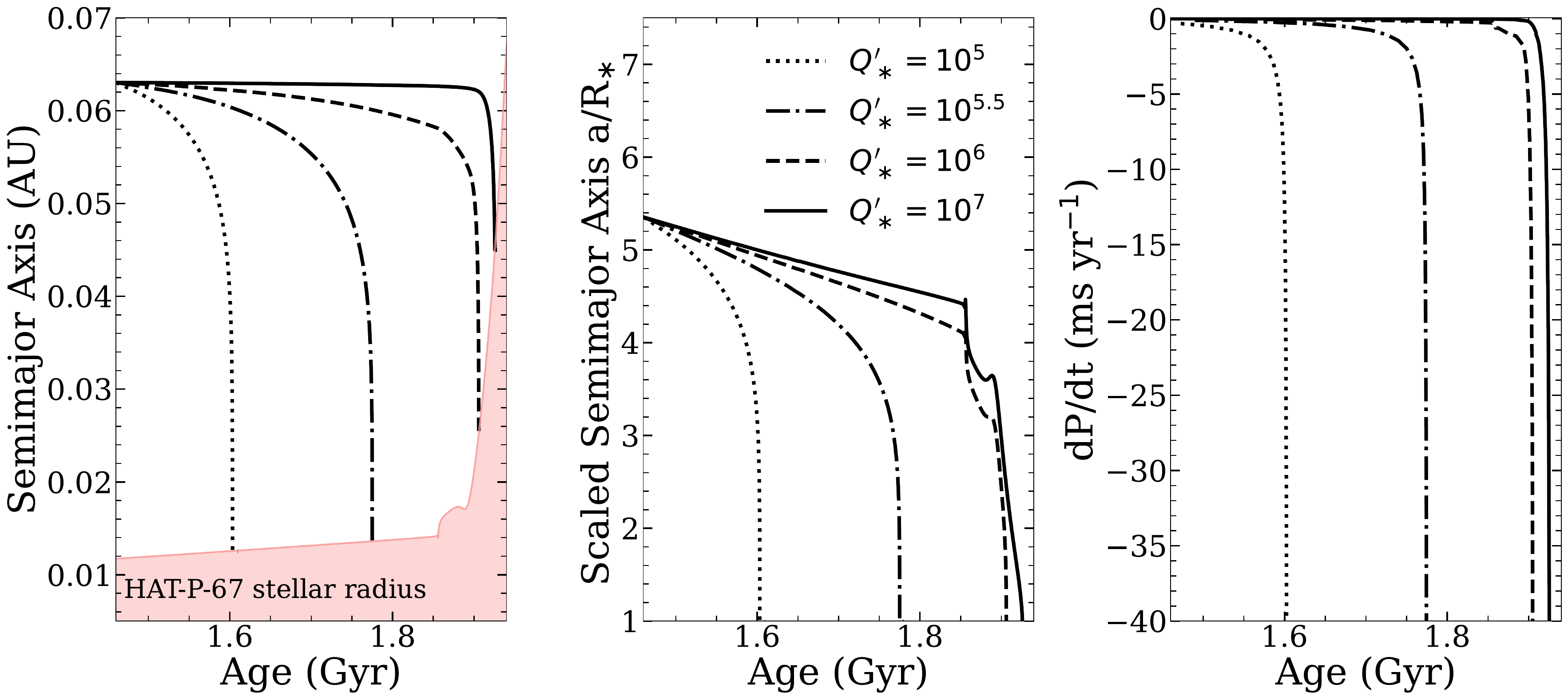}
    \caption{Model for the self-consistent tidal evolution of the HAT-P-67 system.
    We plot as black lines the semimajor axis (left), scaled semimajor axis $a/R_{\ast}$ (center), and orbital decay rate $dP/dt$ (right)  as functions of time in Gyr. 
    As the tidal quality factor $Q_{\ast}'$ for the host star is unknown, we model four potential cases spanning the plausible range of hot Jupiter host stars \citep{Bar20, Wei24}: $10^5$ (dotted lines), $10^{5.5}$ (dash-dotted lines), $10^6$ (dashed lines), and $10^7$ (solid lines).
    We start the planet at a semimajor axis of 0.06303~AU (corresponding to its present-day period of 4.8~days), assuming $Q_{\text{p}} = 10^6$, a circular orbit, synchronized rotation and revolution, and a small nonzero obliquity of 2.2$^\circ$ \citep{Sic24}.
    We evolve the planet from its current age of 1.46 Gyr to the point it reaches its host star's radius.
    If HAT-P-67's $Q_{\ast}'$ is on the higher end of the typical range, the planet will be engulfed in about 500 Myr; on the other hand, if HAT-P-67's $Q_{\ast}'$ is on the lower end of the typical range similar to that of WASP-12 \citep{Yee20}, then the planet will be tidally disrupted in 150 to 250 Myr, demonstrating HAT-P-67~b to be a doomed world. 
    Regardless of which scenario occurs, this orbital decay is unlikely to be detectable for any plausible $Q_{\ast}'$ value.
    }
    \label{fig:tidal_evolution}
\end{figure*}

To investigate potential futures of HAT-P-67~b and whether orbital decay may be observable, we integrate the coupled system of ordinary differential equations (ODEs) modeling the tidal evolution process presented in \citet{Lec10}, following a similar process as \citet{Sch24}.
This system of ODEs is based on the complete tidal evolution equations of the \citet{Hut81} model and is valid at any order in eccentricity, obliquity, and spin.
We use the MESA stellar track radius and the newly derived planet mass and radius as our planet parameter inputs.
We account for the rotation evolution of the host star due to stellar evolution by calculating its current angular momentum with our TESS-derived rotation period.
We use the stellar moment of inertia coefficient from a solar-composition, median-rotation $M_{\ast} = 1.5~M_{\odot}$ stellar model from the \citet{Ama19} grid.
As the star expands, we slow the star's rotation to conserve angular momentum.
We assume that the planet is on a circular orbit, that its rotation period is synchronized with its orbital period, that its tidal quality factor $Q_{\text{p}} = 10^6$, and that its present-day obliquity matches its slight misalignment of 2.2$^{\circ}$ \citep{Sic24}.
We vary the stellar tidal quality factor $Q_{\ast}'$ across a plausible range of values for hot Jupiters \citep{Bar20, Wei24}.
We find $Q_{\ast}' \lesssim 10^6$ would result in the planet's tidal disruption in $150-250$~Myr, prior to the star's rapid expansion as it evolves off the main sequence, while $Q_{\ast}' \gtrsim 10^6$ would result in the planet's engulfment by the star in about 500~Myr, during the star's post-main sequence evolution. We conclude that 500~Myr is an upper bound for the disruption/engulfment timescale. In both scenarios, it is unlikely for any orbital decay to be observable ($|dP/dt| < 1~\mathrm{ms~yr^{-1}} $) at the present day. 
We plot the result of this calculation in Figure \ref{fig:tidal_evolution}.

\section{Conclusion} \label{sec:conclusion}

In this work, we present an analysis of the HAT-P-67~b system, making use of existing ground- and space-based photometry in addition to newly acquired RV data. We update the various parameters of HAT-P-67~b to greater precision, and in particular update its mass to be $0.45 \pm 0.15$~M\textsubscript{J}, consistent with previous measurements but with a uncertainty which has been reduced from 75\% \citepalias{Zhou_2017} to 33\%. 

Additionally, our work underscores the impact stellar activity may have on the determination of planet parameters. Through a detailed analysis of the available photometric and spectroscopic data, we find that the rapid rotation of the host star (and potentially SPMI) induces radial velocity signals at timescales comparable to the orbital period of the planet and with amplitudes comparable to the planetary amplitude. We use a physically motivated quasi-periodic Gaussian process to disentangle the stellar activity from the star's barycentric motion, which yields a mass measurement that is consistent with using no Gaussian process but has a more realistic uncertainty that captures the RV variability due to the stellar activity (see \S\ref{subsubsec:comparison}). 

Our work solidifies HAT-P-67~b as one of the lowest-density transiting planets known to date, as one of only four hot ($T_\mathrm{eq}>1000~\mathrm{K}$) planets with densities below $0.1 \mathrm{~g} \mathrm{~cm}^{-3}$. Models of runaway mass loss imply that HAT-P-67~b will quickly evaporate on the timescale of $\sim0.05\,\mathrm{Gyr}$ \citep{Thorngren_2023}, leaving only the planet's core (a few to tens of Earth masses) behind. HAT-P-67~b's unique position in the hot-Saturn population is due to 1) the recent evolution of the host star off the main sequence, which has not had enough time to strip away the planet's atmosphere and engulf its core, and 2) fortuitous timing, allowing us to capture the planet in its current highly inflated state. 

HAT-P-67~b's low density, now measured to 33\% precision, positions it as a unique target for atmospheric characterization through transmission spectroscopy. There are already multiple existing works in the literature which find a large escaping helium tail and high mass loss rates for this planet \citep{gully_2023, Bello_Arufe_2023, Sic24}, and the bulk parameter constraints that we have placed on HAT-P-67~b will aid in the interpretation of both past and future observations of this intriguing target. 

\section*{Acknowledgments}

\par We would like to thank the anonymous reviewer for their feedback, which improved this paper. We would like to thank the firefighters for their bravery and service to the observatory in the face of the Contreras wildfire. We would like to sincerely thank David Ardila, Heidi Schweiker, and the WIYN board and management for their generosity in managing our program in the aftermath of the Contreras wildfire.

\par This work uses data collected from the WIYN telescope situated on I’oligam Du’ag (Kitt Peak). The authors would like to extend their deepest gratitude to the Tohono O'odham Nation for their stewardship of Kitt Peak and recognize the important cultural significance this mountain holds for the indigenous community. The authors are most fortunate to conduct astronomical observations from this land. 

\par This paper contains data taken with the NEID instrument, which was funded by the NASA-NSF Exoplanet Observational Research (NN-EXPLORE) partnership and built by the Pennsylvania State University. NEID is installed on the WIYN telescope, which is operated by the National Optical Astronomy Observatory, and the NEID archive is operated by the NASA Exoplanet Science Institute at the California Institute of Technology. NN-EXPLORE is managed by the Jet Propulsion Laboratory, California Institute of Technology under contract with the National Aeronautics and Space Administration. This work was supported by a NASA WIYN PI Data Award, administered by the NASA Exoplanet Science Institute.

\par Data presented in this paper are based on observations obtained with the Hungarian-made Automated Telescope Network at the HAT station at the Submillimeter Array of SAO and the HAT station at the Fred Lawrence Whipple Observatory of SAO. 

\par This work is based on observations obtained with the 1.2 m telescope operated by the Smithsonian Astrophysical Observatory at Fred Lawrence Whipple Observatory in Arizona. 

\par This research has made use of the VizieR catalogue access tool, CDS, Strasbourg, France (DOI: 10.26093/
cds/vizier). The original description of the VizieR service was published in \citet{vizier}.

\par This research has made use of the NASA Exoplanet Archive, which is operated by the California Institute of Technology, under contract with the National Aeronautics and Space Administration under the Exoplanet Exploration Program.

\par This paper includes data collected with the TESS mission, obtained from the MAST data archive at the Space Telescope Science Institute (STScI). The TESS data used in this work can be found at\dataset[10.17909/4c8f-f930]{https://dx.doi.org/10.17909/4c8f-f930}. Funding for the TESS mission is provided by the NASA Explorer Program. STScI is operated by the Association of Universities for Research in Astronomy, Inc., under NASA contract NAS 5–26555.  We acknowledge the use of public TESS data from pipelines at the TESS Science Office and at the TESS Science Processing Operations Center.

\par This work has made use of data from the European Space Agency (ESA) mission Gaia (\url{https://www.cosmos.esa.int/gaia}), processed by the Gaia Data Processing and Analysis Consortium (DPAC, \url{https://www.cosmos.esa.int/web/gaia/dpac/consortium}). Funding for the DPAC has been provided by national institutions, in particular the institutions participating in the Gaia Multilateral Agreement.

\par This publication makes use of data products from the Two Micron All Sky Survey, which is a joint project of the University of Massachusetts and the Infrared Processing and Analysis Center/California Institute of Technology, funded by the National Aeronautics and Space Administration and the National Science Foundation.

\par This publication makes use of data products from the Wide-field Infrared Survey Explorer, which is a joint project of the University of California, Los Angeles, and the Jet Propulsion Laboratory/California Institute of Technology, funded by the National Aeronautics and Space Administration. 

\par G.W. would like to thank Samuel Grunblatt, Mercedes L\'opez-Morales, and N\'estor Espinoza for helpful discussions. 

\facilities{ADS, CDS, CTIO:2MASS, Exoplanet Archive, FLWO:2MASS, FLWO:1.2m, Gaia, GALEX, HATNet, IRSA, MAST, NEOWISE, TESS, WISE, WIYN}

\software{\texttt{juliet} \citep{Espinoza2019}, \texttt{astropy} \citep{2013A&A...558A..33A,2018AJ....156..123A, Astropy_2022}, \texttt{dynesty} \citep{Speagle2020}, \texttt{isochrones} \citep{mor15}, \texttt{R} \citep{r24}, \texttt{matplotlib} \citep{Hunter_2007}, \texttt{numpy} \citep{Harris_2020}, \texttt{scipy} \citep{2020SciPy-NMeth}, \texttt{pandas} \citep{mckinney-proc-scipy-2010, reback2020pandas}}

\appendix

\setcounter{table}{0} 
\renewcommand{\thetable}{A\arabic{table}}
\renewcommand\theHtable{Appendix.\thetable}

\setcounter{figure}{0}
\renewcommand{\thefigure}{A\arabic{figure}}
\renewcommand\theHfigure{Appendix.\thefigure}

\section{Additional Tables and Plots}

\startlongtable
\begin{longrotatetable}
\setlength{\tabcolsep}{10pt}
\begin{deluxetable*}{ccccccccccc}
\tablewidth{0pt}
\tablecaption{First 20 lines of activity indicator data from NEID. The full table is available online in machine-readable form.} \label{tab:indicators_extended_1}
\tablehead{
\colhead{BJD\textsubscript{TDB}} & \colhead{dLW} & \colhead{$\sigma$\textsubscript{dLW}} & \colhead{CRX} & \colhead{$\sigma$\textsubscript{CRX}} & \colhead{CaIRT 1} & \colhead{$\sigma$\textsubscript{CaIRT 1}} & \colhead{CaIRT 2} & \colhead{$\sigma$\textsubscript{CaIRT 2}} & \colhead{CaIRT 3} & \colhead{$\sigma$\textsubscript{CaIRT 3}}
}
\startdata
$2459723.87646187$ & $-2563.23$ & $847.30$ & $375.36$ & $380.59$ & $0.6111$ & $0.0056$ & $0.4425$ & $0.0061$ & $0.4793$ & $0.0065$ \\
$2459723.88356815$ & $-3590.18$ & $651.13$ & $363.27$ & $310.55$ & $0.6144$ & $0.0044$ & $0.4576$ & $0.0046$ & $0.4838$ & $0.0050$ \\
$2459723.89063798$ & $-2501.34$ & $604.01$ & $-75.14$ & $304.31$ & $0.6069$ & $0.0042$ & $0.4620$ & $0.0044$ & $0.4824$ & $0.0047$ \\
$2459725.90326585$ & $1683.63$ & $260.33$ & $-198.65$ & $116.69$ & $0.6170$ & $0.0021$ & $0.4547$ & $0.0020$ & $0.4899$ & $0.0022$ \\
$2459725.91048804$ & $2004.25$ & $255.93$ & $-39.87$ & $105.71$ & $0.6164$ & $0.0020$ & $0.4532$ & $0.0020$ & $0.4877$ & $0.0021$ \\
$2459725.91775956$ & $2416.96$ & $277.84$ & $-43.98$ & $120.15$ & $0.6163$ & $0.0022$ & $0.4555$ & $0.0021$ & $0.4865$ & $0.0023$ \\
$2459728.76095795$ & $221.40$ & $387.00$ & $-77.90$ & $179.63$ & $0.6192$ & $0.0028$ & $0.4573$ & $0.0028$ & $0.4905$ & $0.0030$ \\
$2459728.76826597$ & $255.22$ & $356.06$ & $-217.52$ & $137.09$ & $0.6193$ & $0.0026$ & $0.4587$ & $0.0027$ & $0.4885$ & $0.0028$ \\
$2459728.77558364$ & $490.23$ & $313.77$ & $34.92$ & $134.28$ & $0.6157$ & $0.0024$ & $0.4563$ & $0.0023$ & $0.4905$ & $0.0025$ \\
$2459732.86293669$ & $-26.59$ & $370.24$ & $101.85$ & $181.34$ & $0.6145$ & $0.0028$ & $0.4622$ & $0.0028$ & $0.4875$ & $0.0030$ \\
$2459732.87008587$ & $583.31$ & $354.66$ & $-131.23$ & $158.26$ & $0.6135$ & $0.0027$ & $0.4605$ & $0.0027$ & $0.4894$ & $0.0029$ \\
$2459732.87739681$ & $31.95$ & $341.16$ & $-176.14$ & $152.08$ & $0.6187$ & $0.0026$ & $0.4610$ & $0.0026$ & $0.4897$ & $0.0027$ \\
$2459733.72969579$ & $-518.96$ & $551.45$ & $-55.40$ & $246.16$ & $0.6059$ & $0.0039$ & $0.4516$ & $0.0040$ & $0.4923$ & $0.0043$ \\
$2459733.73686593$ & $-1203.55$ & $579.72$ & $-113.71$ & $252.05$ & $0.6156$ & $0.0042$ & $0.4625$ & $0.0043$ & $0.4850$ & $0.0046$ \\
$2459733.74416046$ & $-201.61$ & $614.12$ & $572.30$ & $273.96$ & $0.6155$ & $0.0044$ & $0.4612$ & $0.0045$ & $0.4931$ & $0.0048$ \\
$2459737.87190599$ & $-11.09$ & $431.02$ & $78.88$ & $185.51$ & $0.6128$ & $0.0031$ & $0.4645$ & $0.0031$ & $0.4913$ & $0.0034$ \\
$2459737.87914954$ & $174.83$ & $387.83$ & $-20.92$ & $172.68$ & $0.6197$ & $0.0028$ & $0.4614$ & $0.0029$ & $0.4897$ & $0.0031$ \\
$2459737.88617648$ & $576.71$ & $373.17$ & $363.04$ & $149.03$ & $0.6148$ & $0.0028$ & $0.4619$ & $0.0028$ & $0.4854$ & $0.0030$ \\
$2459738.65439909$ & $0.63$ & $382.38$ & $32.86$ & $163.78$ & $0.6198$ & $0.0027$ & $0.4576$ & $0.0027$ & $0.4888$ & $0.0029$ \\
$2459738.66170780$ & $-479.77$ & $370.98$ & $-80.75$ & $173.98$ & $0.6194$ & $0.0027$ & $0.4578$ & $0.0027$ & $0.4922$ & $0.0028$ \\
\enddata
\end{deluxetable*}
\end{longrotatetable}

\startlongtable
\begin{longrotatetable}
\setlength{\tabcolsep}{12pt}
\begin{deluxetable*}{ccccccccccc}
\tablewidth{0pt}
\tablecaption{Continued from Table \ref{tab:indicators_extended_1}. The full table is available online in machine-readable form.} \label{tab:indicators_extended_2}
\tablehead{
\colhead{H$\alpha$} & \colhead{$\sigma$\textsubscript{H$\alpha$}} & \colhead{NaD 1} & \colhead{$\sigma$\textsubscript{NaD 1}} & \colhead{NaD 2} & \colhead{$\sigma$\textsubscript{NaD 2}} & \colhead{PaD} & \colhead{$\sigma$\textsubscript{PaD}} & \colhead{NaNIR} & \colhead{$\sigma$\textsubscript{NaNIR}} & \colhead{Instrument$^{\text{a}}$}
}
\startdata
$0.3885$ & $0.0035$ & $0.6623$ & $0.0054$ & $0.7366$ & $0.0056$ & $0.694$ & $0.069$ & $0.0518$ & $0.0061$ & NEID1 \\
$0.3857$ & $0.0027$ & $0.6643$ & $0.0042$ & $0.7430$ & $0.0044$ & $0.763$ & $0.050$ & $0.0530$ & $0.0048$ & NEID1 \\
$0.3862$ & $0.0025$ & $0.6724$ & $0.0040$ & $0.7357$ & $0.0041$ & $0.710$ & $0.045$ & $0.0500$ & $0.0046$ & NEID1 \\
$0.3876$ & $0.0011$ & $0.6659$ & $0.0019$ & $0.7397$ & $0.0020$ & $0.762$ & $0.018$ & $0.0528$ & $0.0022$ & NEID1 \\
$0.3870$ & $0.0011$ & $0.6686$ & $0.0018$ & $0.7440$ & $0.0019$ & $0.745$ & $0.017$ & $0.0513$ & $0.0022$ & NEID1 \\
$0.3875$ & $0.0012$ & $0.6634$ & $0.0020$ & $0.7387$ & $0.0021$ & $0.745$ & $0.019$ & $0.0539$ & $0.0024$ & NEID1 \\
$0.3889$ & $0.0016$ & $0.6578$ & $0.0026$ & $0.7375$ & $0.0028$ & $0.711$ & $0.026$ & $0.0546$ & $0.0031$ & NEID1 \\
$0.3875$ & $0.0015$ & $0.6605$ & $0.0024$ & $0.7436$ & $0.0026$ & $0.708$ & $0.025$ & $0.0540$ & $0.0029$ & NEID1 \\
$0.3872$ & $0.0013$ & $0.6612$ & $0.0022$ & $0.7456$ & $0.0023$ & $0.735$ & $0.021$ & $0.0509$ & $0.0026$ & NEID1 \\
$0.3875$ & $0.0016$ & $0.6744$ & $0.0026$ & $0.7316$ & $0.0027$ & $0.700$ & $0.026$ & $0.0509$ & $0.0030$ & NEID1 \\
$0.3888$ & $0.0015$ & $0.6682$ & $0.0025$ & $0.7364$ & $0.0026$ & $0.724$ & $0.024$ & $0.0469$ & $0.0029$ & NEID1 \\
$0.3870$ & $0.0014$ & $0.6717$ & $0.0024$ & $0.7354$ & $0.0025$ & $0.719$ & $0.022$ & $0.0478$ & $0.0028$ & NEID1 \\
$0.3887$ & $0.0023$ & $0.6705$ & $0.0037$ & $0.7471$ & $0.0038$ & $0.745$ & $0.042$ & $0.0563$ & $0.0042$ & NEID1 \\
$0.3888$ & $0.0024$ & $0.6707$ & $0.0038$ & $0.7394$ & $0.0040$ & $0.814$ & $0.045$ & $0.0498$ & $0.0044$ & NEID1 \\
$0.3912$ & $0.0026$ & $0.6589$ & $0.0040$ & $0.7432$ & $0.0042$ & $0.668$ & $0.046$ & $0.0532$ & $0.0046$ & NEID1 \\
$0.3876$ & $0.0018$ & $0.6630$ & $0.0029$ & $0.7488$ & $0.0031$ & $0.740$ & $0.031$ & $0.0430$ & $0.0034$ & NEID1 \\
$0.3893$ & $0.0016$ & $0.6637$ & $0.0026$ & $0.7512$ & $0.0028$ & $0.745$ & $0.028$ & $0.0412$ & $0.0031$ & NEID1 \\
$0.3882$ & $0.0015$ & $0.6651$ & $0.0026$ & $0.7476$ & $0.0027$ & $0.738$ & $0.027$ & $0.0432$ & $0.0030$ & NEID1 \\
$0.3913$ & $0.0015$ & $0.6600$ & $0.0026$ & $0.7467$ & $0.0027$ & $0.666$ & $0.024$ & $0.0406$ & $0.0030$ & NEID1 \\
$0.3910$ & $0.0015$ & $0.6598$ & $0.0025$ & $0.7470$ & $0.0027$ & $0.704$ & $0.024$ & $0.0468$ & $0.0029$ & NEID1 \\
\enddata
\begin{itemize}
    \item[] $^{\text{a}}$ NEID1 refers to data collected before the June 2022 Contreras Fire; NEID2 refers to data collected after 
\end{itemize}
\end{deluxetable*}
\end{longrotatetable}

\begin{figure*}[ht!]
    \centering
    \includegraphics[width=\textwidth]{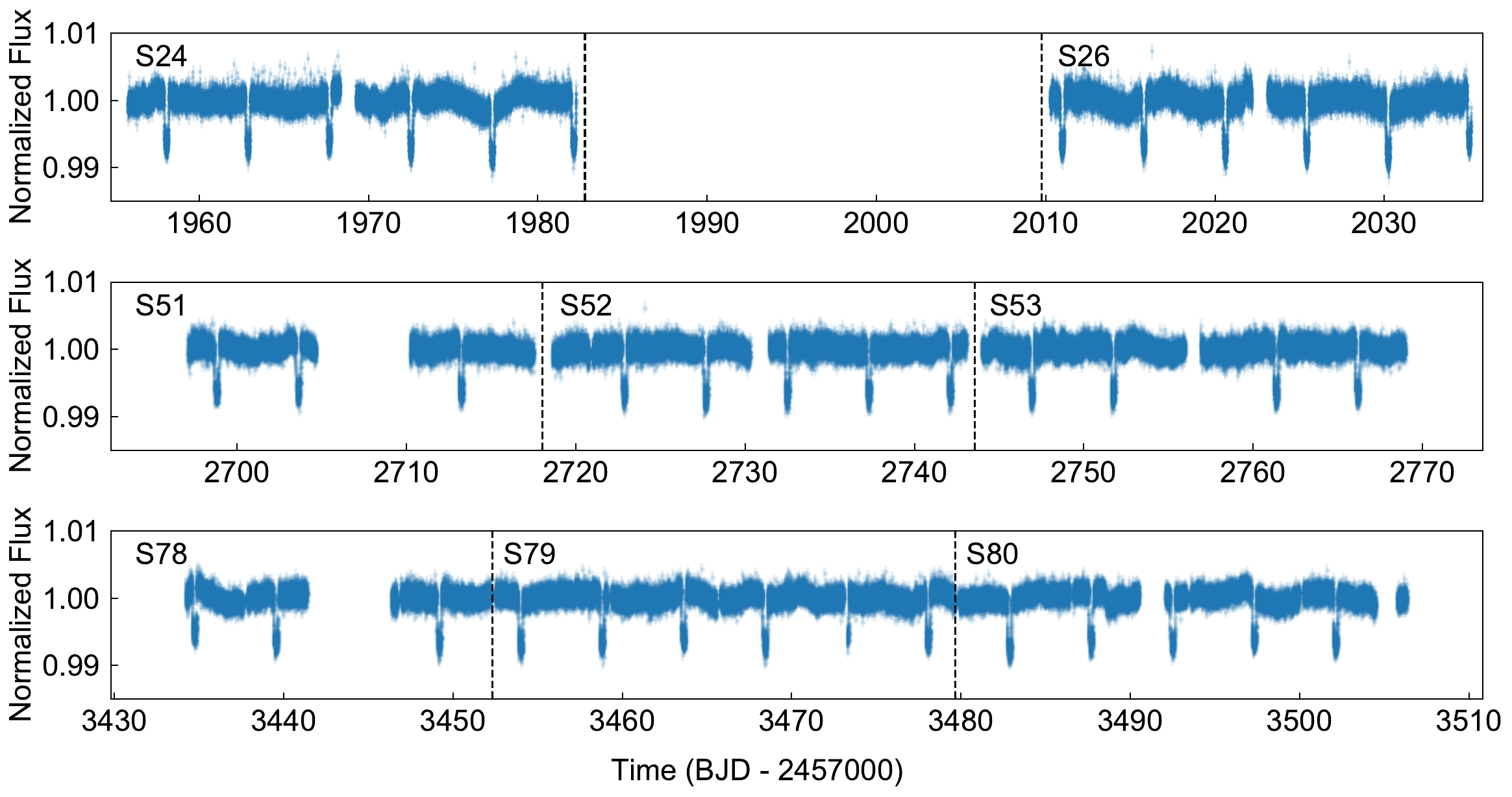}
    \caption{Full light curves from TESS. Data from Years 2, 4, and 6 of the TESS mission are shown in the top, middle, and bottom rows respectively. The data visually suggest a stellar rotation period of a few days. Stellar activity is strongest in Year 2, weakens for Year 4, and returns to elevated levels in Year 6, demonstrating starspots+rotation as a probable cause of the variations.}
    \label{fig:oot}
\end{figure*}

\begin{figure*}
    \centering
    \includegraphics[width=\textwidth]{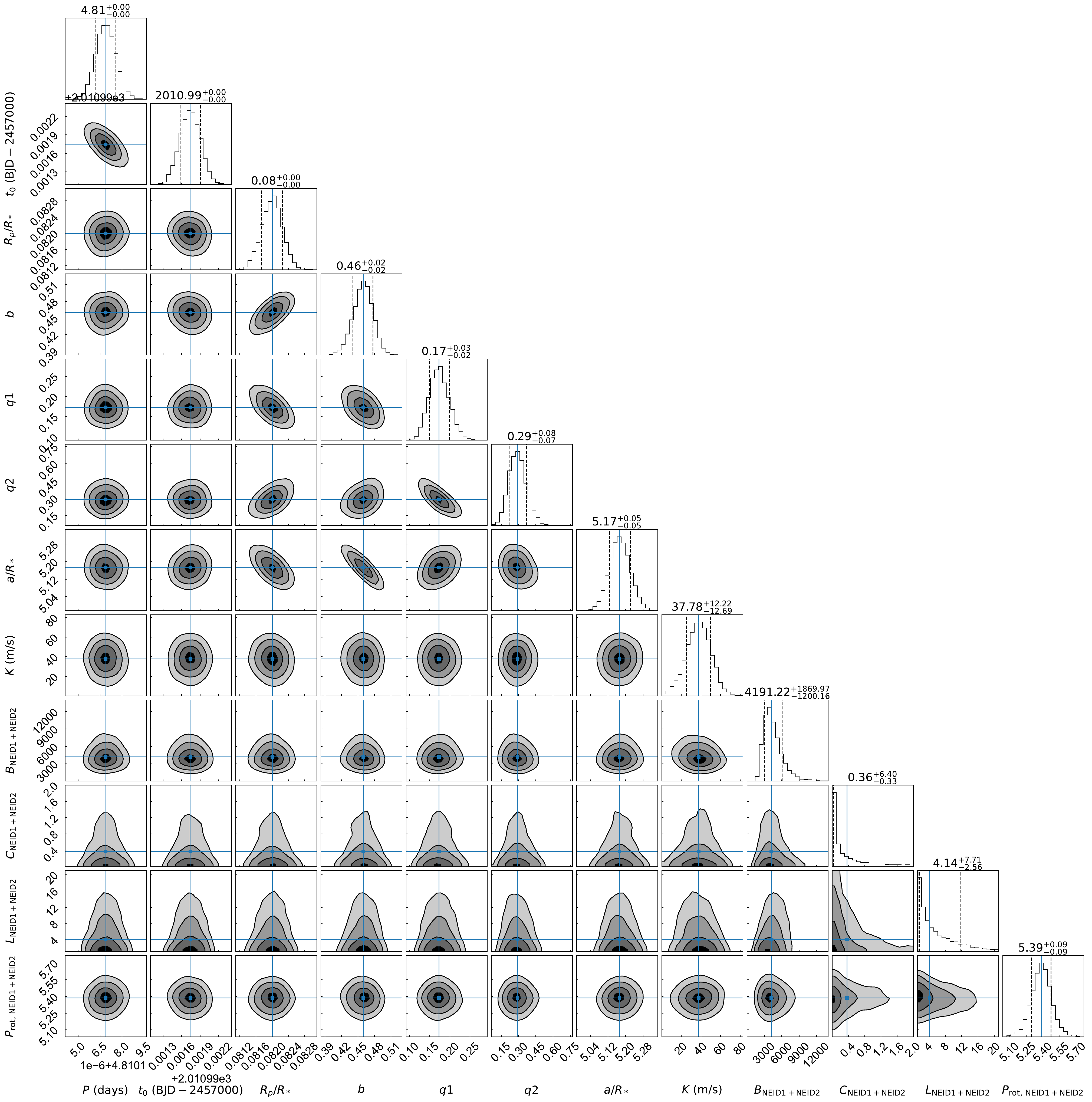}
    \caption{Corner plot for joint (photometry+RVs) fit which includes one planet and a quasi-periodic GP. We adopt this fit as final. We include only the planetary parameters and GP parameters for the NEID data for clarity.}
    \label{fig:corner}
\end{figure*}

\begin{figure*}
    \centering
    \includegraphics[width=\linewidth]{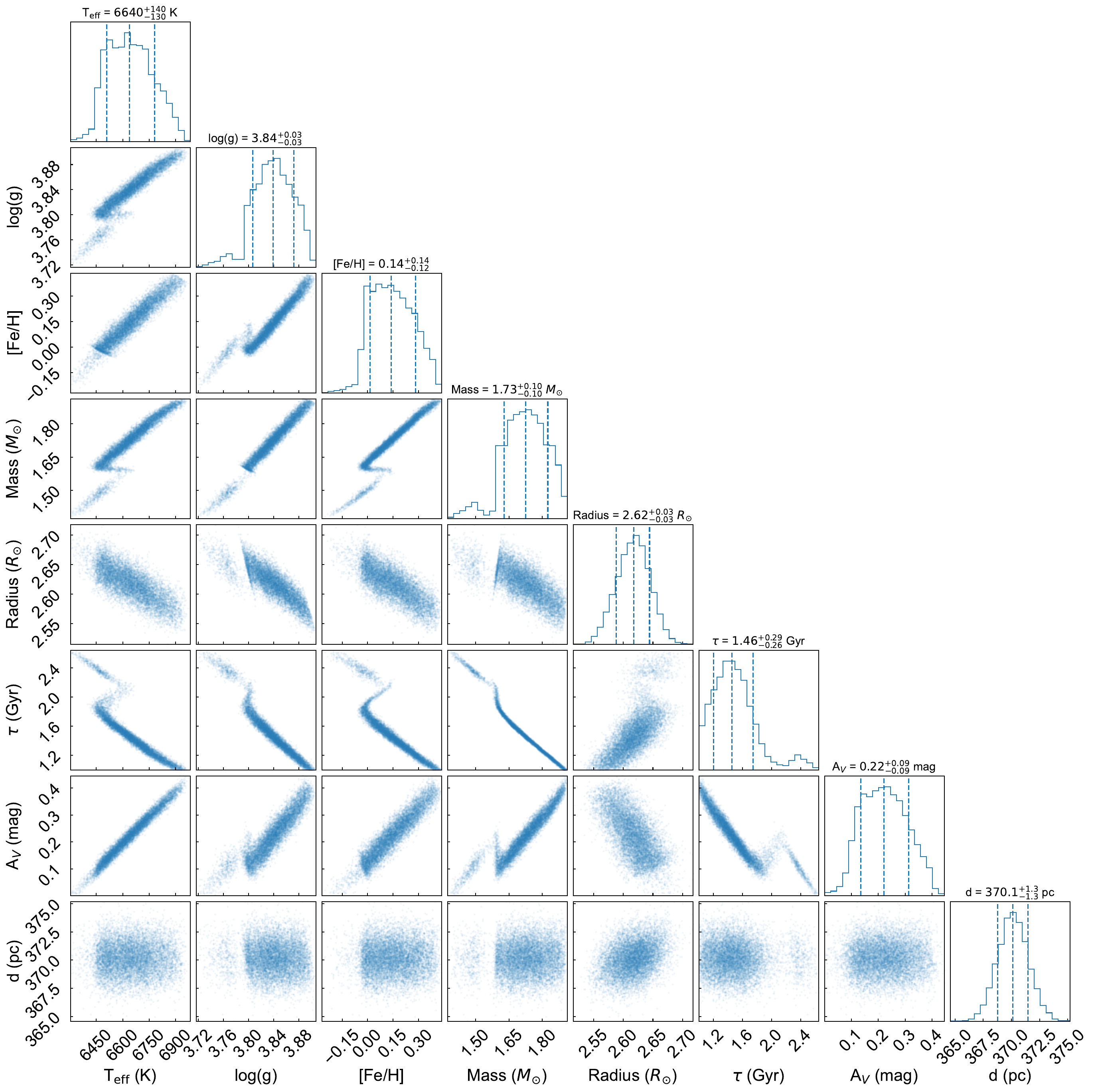}
    \caption{Corner plot for HAT-P-67A fit.}
    \label{fig:host_corner}
\end{figure*}

\bibliography{neid_hatp67b}{}
\bibliographystyle{aasjournal}

\end{document}